\documentclass[aps,prd,superscriptaddress,twocolumn,10pt,amsmath,amssymb,nofootinbib]{revtex4-1}
\usepackage{acronym}
\usepackage{graphics}
\usepackage{graphicx}
\usepackage{subfigure}
\usepackage{siunitx}
\usepackage[usenames]{color}
\usepackage{xcolor}
\usepackage{url}
\usepackage[normalem]{ulem}
\DeclareRobustCommand{\erase}{\bgroup\markoverwith{\textcolor{blue}{\rule[.5ex]{2pt}{0.4pt}}}\ULon}

\newcommand{\abs}[1]{\left\lvert #1 \right\rvert}
\newcommand{\av}[1]{\left\langle #1 \right\rangle}
\newcommand{\Omegagw}{\Omega_{\textrm{GW}}}

\newcommand{\figref}[1]{Fig.\@~\ref{#1}}

\begin{document}

%%%%%%%%%%%%%%%%%%%%%%%%%%%%%%%%%%%%%%%%%%%%%%%%%%%%%%%%%%%%%%%%%%%%%%%%%%%%%%%
% TITLE, AUTHOR
%%%%%%%%%%%%%%%%%%%%%%%%%%%%%%%%%%%%%%%%%%%%%%%%%%%%%%%%%%%%%%%%%%%%%%%%%%%%%%%

\title{Applicability of multi-component study on Bayesian searches for targeted anisotropic stochastic gravitational-wave background}

\author{Soichiro Kuwahara}
\affiliation{Research Center for the Early Universe (RESCEU), The University of Tokyo, Tokyo 113-0033, Japan}
\author{Leo Tsukada}
\affiliation{Department of Physics and Astronomy, University of Nevada, Las Vegas, 4505 South Maryland Parkway, Las Vegas, NV 89154, USA}
\affiliation{Nevada Center for Astrophysics, University of Nevada, Las Vegas, NV 89154, USA}

\begin{abstract}
Stochastic background gravitational waves have not yet been detected by ground-based laser interferometric detectors, but recent improvements in detector sensitivity have raised considerable expectations for their eventual detection. Previous studies have introduced methods for exploring anisotropic background gravitational waves using Bayesian statistics. These studies represent a groundbreaking approach by offering physically motivated anisotropy mapping that is distinct from the Singular Value Decomposition regularization of the Fisher Information Matrix. However, they are limited by the use of a single model, which can introduce potential bias when dealing with complex data that may consist of a mixture of multiple models. Here, we demonstrate the bias introduced by a single-component model approach in the parametric interpretation of anisotropic stochastic gravitational-wave backgrounds, and we confirm that using multiple-component models can mitigate this bias.
\end{abstract}

%\pacs{\FIXME{FIXME}}
\maketitle
\acrodef{CSD}{cross spectral density}
\acrodef{GW}{gravitational-wave}
\acrodef{SNR}{signal-to-noise ratio}
\acrodef{PDF}{probability density function}
\acrodef{PSD}{power spectral density}
\acrodef{SGWB}{stochastic gravitational-wave background}
\acrodef{CBC}{compact binary coalescence}
\acrodef{BBH}{binary black hole}
\acrodef{BNS}{binary neutron star}
\acrodef{PBH}{primordial black holes}
\acrodef{PTA}{pulsar timing array}
\acrodef{SPH}{spherical harmonics}
\acrodef{BBR}{broadband radiometer analysis}
\acrodef{BF}{Bayes factor}
\acrodef{NBR}{narrow band radiometer analysis}
\acrodef{LVK}{LIGO, Virgo and KAGRA}

%%%%%%%%%%%%%%%%%%%%%%
\section{Introduction}
The \ac{SGWB} is a super-position of numerous gravitational waves whose sources are not individually resolved. Several astrophysical and cosmological processes have been proposed as contributing to the \ac{SGWB}. The astrophysical sources include, for instance, large numbers of distant \acp{BBH}, \acp{BNS} \cite{Regimbau2008,Rosado2011, Wu2012,Zhu2013,Zhu_2011,Marassi_2011}, magnetars \cite{Regimbau_2001,Marassi_2010,Wu_2013}, and core-collapse supernovae \cite{Buonanno2005,Howell_2004,Sandick_2006,Marassi_2009,Zhu_2010}. On the other hand, cosmological sources include cosmic strings \cite{Sarangi_2002,Damour_2005,Siemens_2007,Kibble_1976}, \acp{GW} emitted during the inflationary era \cite{Starobinsky_1979,Turner_1997,Bar_1994}, and \acp{PBH} \cite{Sasaki_2016,Mandic_2016,Wang_2018}.

Recently, the Advanced LIGO (aLIGO) \cite{DetectorALIGO} and Advanced Virgo \cite{DetectorAVirgo} detectors have completed the third observing run (O3) and the first period of the fourth observing run (O4a), collaborating with KAGRA \cite{DetectorKAGRA}. The ground-based \ac{GW} detectors have not made any detection of the \ac{SGWB} \cite{O3Isotropic}. However, in the nanohertz frequency band, where ground-based \ac{GW} detectors are not sensitive due to seismic noise, \ac{PTA} collaborations have reported a potential signal in the form of a \ac{SGWB}-induced quadrupolar correlation of timing residuals \cite{CPTA2023Jun,nanoGrav2023Jun,EPTA2023Jun,PPTA2023Jun}.

The promising results from \ac{PTA} collaborations emphasize the importance of searching for not only isotropic but also anisotropic \ac{SGWB}. Some models of the aforementioned astrophysical or cosmological sources imply characteristic angular distributions \cite{Contaldi_2017,Jenkins_2019_oct,Jenkins_2019_Sep,Bertacca_2020,Cusin_2017,Cusin_2018,Cusin_2018_2,Cusin_2019,Pitrou_2020,Ca_as_Herrera_2020,Geller_2018}. Furthermore, persistent point sources such as Scorpius X-1 \cite{Abbott_2017,Abbott_2017_Jun} might also be observable by the detection techniques developed for anisotropic \ac{SGWB}. The \ac{LVK} collaboration reported the results of their search for anisotropic \ac{SGWB} \cite{O3Anisotropic}. They introduced three analysis techniques: \ac{BBR} \cite{BBR_LIGO, BBR_Ballmer}, \ac{NBR} \cite{NBR_Ballmer}, and \ac{SPH} \cite{Sph_Thrane} analysis, derived from cross-correlation techniques and the choice of orthonormal basis.

Inspired by \ac{SPH} analysis, Tsukada \textit{et al.}\@ \cite{Tsukada2023} presented a Bayesian parameter estimation approach for a targeted anisotropic \ac{SGWB}. The conventional \ac{SPH} analysis approach, which was applied in the \ac{LVK} collaboration search \cite{O3Anisotropic}, estimates each spherical-harmonic mode using regularization techniques. On the other hand, Tsukada \textit{et al.}\@ \cite{Tsukada2023} formalized the analysis assuming that a specific anisotropy model was known and estimated its parameters while the search is limited to employing a single model. We extend the previous Bayesian targeted anisotropy analysis \cite{Tsukada2023} to the multi-template inference framework and quantify the resulting biases and detection thresholds for realistic sky patterns. Although there are several searches for multi-component isotropic-only or anisotropic-only \ac{SGWB} \cite{Parida_2016,Boileau_2021,Suresh_2021,Federico_2024} that try to separate the signal components by assuming different frequency dependencies, there have been no attempts to separate signals by assuming differences in their anisotropic features. In this work, we introduce the component separation technique which separates the signals in a multi-component anisotropic \ac{SGWB} based on the $\mathcal{P}_{lm}$ distributions of the components.

We also demonstrate the bias because we did not consider the isotropic component in the search for anisotropy.
Some of the methods currently applied in the \ac{LVK} studies could be affected by this bias. For example, Refs.~\cite{BBR_LIGO,BBR_Ballmer,NBR_Ballmer} perform the pixel-based analysis under the assumption that a point-like anisotropic source is louder than the isotropic \ac{SGWB}, allowing for neglecting correlations between the intensities of neighboring pixels. However, this assumption may no longer hold when the isotropic signal becomes comparable to the anisotropic sources, which can occur in regimes with improved detector sensitivity. Although the quantitative analysis of the bias on those methods shall be discussed separately from this search and remain a subject for future work.
Although our technique is still limited to \textit{a priori} choices of anisotropic distribution models, we obtain the clean map with constraints motivated by physical processes. 

In this paper, we assume a physical model consisting of an isotropic component together with a galactic-plane-motivated anisotropic component and demonstrate the bias caused by choosing a single component recovery model. We demonstrate that applying a multi-component recovery model mitigates the bias.  Using probability-probability plots generated from the posterior distributions, we prove the necessity of introducing a multi-component analysis in this Bayesian approach. The main purpose of this paper is to verify the validity of this map separation technique and the risk of bias when choosing a wrong constraint for the sky, which is shown in Sec.\@\ref{sec:ParamEstimation}. Sec.\@\ref{sec:BFanalysis} are our additional studies: we examine changes in \ac{BF} by altering the strength of each signal or altering noise realizations.

\section{Formalism}
\subsection{Likelihood}
The basic Bayesian formalism idea for an anisotropic \ac{SGWB} is stated in Ref.\@ \cite{Tsukada2023}. Furthermore, Tsukada \cite{Tsukada2023nonGR} expanded the likelihood formalism for the presence of the mixed polarization signal. This formalism can be used for multi-component models. We can write the expectation value of the \ac{CSD} as
\begin{align}
    \av{C(f,t)} =\sum_i \sum_{\mu} \gamma_{\mu}(f,t)\mathcal{P}_{\mu}^i(f,\vec{\theta})
\end{align}
where $i$ represents each signal components and $\mu$ represents $(l,m)$ basis since we choose spherical-harmonics basis to decompose the skymap in this search. The definition of \ac{CSD} is the cross-correlated data from two detectors at the frequency $f$ and the time coordinate $t$, and its formulation is given below:
\begin{align}
    C(f,t) \equiv \frac{2}{\tau}\tilde{d}_1(f,t)\tilde{d}_2^*(f,t)
\end{align}
where the $\tilde{d}(f,t)$ is the Fourier transformation of detector output $d(t)$ computed in an interval $[t-\tau/2,t+\tau/2]$. $\tilde{d}(f,t)$ can be expressed by the signal component $\tilde{h}(f,t)$ and noise component $\tilde{n}(f,t)$,
\begin{align}
    \tilde{d}(f,t)=\tilde{h}(f,t)+\tilde{n}(f,t).
\end{align}
$\gamma_{\mu}$ is an overlap function of the detector network for the basis $\mu$ \cite{Allen_ORF} and its multiplication appears as the detector response of the signal. In the weak signal approximation and assuming no cross-correlated noise between detectors, the covariance matrix of \ac{CSD} can be approximated as a diagonal matrix \cite{Romano_2017} given by
%\begin{align}
%    \mathbf{\Sigma}_{CC'}&\equiv\av{\abs{C(f,t)}^2}-%\abs{\av{C(f,t)}}^2\\\nonumber
%    &\approx\delta_{tt'}\delta_{ff'}P_1(f,t)P_2(f,t)
%\end{align}
%where $P_k(f,t)$ is the \ac{PSD} of the $k$-th detector centered around $t$.
\begin{align}
    \mathbf{\Sigma}_{CC'}&\equiv\av{\abs{C(f,t)}^2}-\abs{\av{C(f,t)}}^2\\\nonumber
    &\approx \frac{4}{\tau^2}\left\{\av{\tilde{n}_1^*(f,t)\tilde{n}_1(f',t')}\av{\tilde{n}_2(f,t)\tilde{n}_2^*(f',t')}\right\}.
\end{align}
%We also obtain
The noise $P_k(f,t)$ is the \ac{PSD} of the $k$-th detector centered around $t$ is 
\begin{align}
%    \frac{2}{\tau}\av{\tilde{d}_1^{*}(f,t)\tilde{d}_1(f',t')}
%    &\approx
\frac{2}{\tau}\av{\tilde{n}_1^*(f,t)\tilde{n}_1(f',t')}
    &=\delta_{tt'}\delta_{ff'}P_1(t,f)
\end{align}
%where $P_k(f,t)$ is the \ac{PSD} of the $k$-th detector centered around $t$. 
Therefore, in the weak signal approximation, we use
\begin{align}
    \mathbf{\Sigma}_{CC'}&\approx\delta_{tt'}\delta_{ff'}P_1(t,f)P_2(f,t)
\end{align}
as the covariance matrix.

According to the factorization given in Ref.\@ \cite{Tsukada2023,Tsukada2023nonGR}, assuming the target anisotropy model can be decomposed into frequency spectrum factor $H(f,\vec{\theta})$ and sky-map factor in spherical-harmonic basis $\mathcal{P}_{lm}$, it follows that
\begin{align}
    \mathcal{P}_{lm}(f,\vec{\theta})=\epsilon\bar{H}(f,\vec{\theta})\bar{\mathcal{P}}_{lm}.
\end{align}
$\bar{H}(f,\vec{\theta})$ and $\bar{\mathcal{P}}_{lm}$ are normalized as
\begin{align}
    \bar{H}(f,\vec{\theta}) & \equiv \frac{H(f,\vec{\theta})}{H(f_{\mathrm{ref}},\vec{\theta})},\\
    \bar{\mathcal{P}}_{lm} & \equiv \frac{\mathcal{P}_{lm}}{\mathcal{P}_{00}}. \label{eq:Plm_nornalized}
\end{align}

In addition, $\epsilon$ is normalized by $\frac{3H_0^2}{2\pi^2f_{\mathrm{ref}}^3\sqrt{4\pi}}$ so that the amplitude parameter $\epsilon$ is comparable with $\hat{\Omega}_{\mathrm{GW}}$ of the isotropic search of \ac{LVK} \cite{O1Isotropic,O2Isotropic,O3Isotropic}. Here, we may use the Gaussian likelihood because, in the actual data analysis, whole observation data (e.g., one year) are folded into one sidereal day \cite{Ain_2015}, which is further divided into short time (e.g., 192 seconds \cite{Ain_2015,Allen_1999}) segments of these folded data. Since typically more than 300 days of data are combined into a single dataset, we can assume the central limit theorem. This method has been applied in recent analyses such as Ref.\@ \cite{O3Anisotropic}. Given multiple signal component $\mathcal{M}$, the form of Gaussian likelihood becomes almost identical to Eq.\@(21) in Ref.\@ \cite{Tsukada2023nonGR}, except that we consider the tensor mode only in this paper: 
\begin{widetext}
\begin{align}
    p({C_{ft}}|\epsilon_i,\vec{\theta_i};\mathcal{M}) = \frac{1}{(2\pi)^{{\rm N}_{\rm dim}/2}\sqrt{|\mathbf{\Sigma}_{CC'}|}}\exp{\left\{-\frac{1}{2}\sum_{f,t}\frac{\tau\Delta f|C(f,t)-\sum_i\epsilon_i\bar{H_i}(f,\vec{\theta})\gamma_{\mu}(f,t)\bar{\mathcal{P}}_{\mu}^i|^2}{P_1(f,t)P_2(f,t)}\right\}}.
    \label{eq:likelihood}
\end{align}
\end{widetext}
The term $\tau\Delta f$ appears because in practice we use discrete data in time and frequency with the segment duration $\tau$ and frequency resolution $\Delta f$. To implement this Bayesian analysis, we used and further developed the pipeline used in \cite{Tsukada2023,Tsukada2023nonGR} which utilizes the \texttt{Bilby} package \cite{bilby_2019, bilby_2020}. Since the integration in a multidimensional parameter space is required for the computation of posterior distribution and evidence as shown in Eq.\@\eqref{eq:likelihood}, we adopt a nested sampling algorithm package called \texttt{Dynesty} \cite{dynesty}.

The calculation of Eq.\@\eqref{eq:likelihood} requires two-dimensional integration over frequencies and time segments, and requires significant computational time. Ref.\@ \cite{Tsukada2023} introduces the implementation of time integration part by precomputing and storing the result beforehand. Furthermore, Ref.\@ \cite{Tsukada2023nonGR} formulates the extra components of the precomputable part in the presence of multiple signal components of various polarization modes. Here, we basically follow the same computation as in Ref.\@ \cite{Tsukada2023nonGR} but only consider the tensor mode with multiple anisotropic signal components. The exponent part of Eq.\@\eqref{eq:likelihood} can be written as
\begin{widetext}
\begin{align}
    &\exp{\left\{-\frac{1}{2}\sum_{f,t}\frac{\tau\Delta f|C(f,t)-\sum_i\epsilon_i\bar{H_i}(f,\vec{\theta})\gamma_{\mu}(f,t)\bar{\mathcal{P}}_{\mu}^i|^2}{P_1(f,t)P_2(f,t)}\right\}}\\\nonumber
    &=\exp{\left\{-\frac{1}{2}\sum_{f,t}\frac{\tau\Delta f|C(f,t)|^2}{P_1(f,t)P_2(f,t)}+\sum_{i}\epsilon_i\Re[{(\bar{\mathcal{P}}_{\mu}^i)^{*}X_{\mu}^i}]-\frac{1}{2}\sum_{i}\sum_{j}\epsilon_i\epsilon_j(\bar{\mathcal{P}}_{\mu}^i)^{*}\Gamma_{\mu\nu}^{ij}\bar{\mathcal{P}}_{\nu}^h\right\}}
\end{align}
\end{widetext}
where the indices $(\mu,\nu)$ corresponds to the summation over spherical-harmonics basis. The definitions of matrix $X_{\mu}^i$ and $\Gamma_{\mu\nu}^{ij}$ are as follows
\begin{align}
    X_{\mu}^i &= \sum_f\bar{H_i}(f,\vec{\theta})\sum_t\frac{\tau\Delta fC(f,t)\gamma^*_{\mu}(f,t)}{P_1(f,t)P_2(f.t)}\\
    \Gamma_{\mu\nu}^{ij} &= \sum_f\bar{H_i}(f,\vec{\theta})\bar{H_j}(f,\vec{\theta'})\sum_t\frac{\tau\Delta f\gamma^*_{\mu}(f,t)\gamma_{\nu}(f,t)}{P_1(f,t)P_2(f,t)}
\end{align}
where $\tau$ is the duration of the segment where the short Fourier transformation is performed to the time series and $\Delta f$ is the frequency resolution. The $\sum_t$ of both equations is precomputed and stored to decrease the computation time.

\section{Parameter Estimation for multi-component injection}
\label{sec:ParamEstimation}

\subsection{Setup}

\subsubsection{$\mathcal{P}_{lm}$ distribution}
\label{subsubsec:Plmmodel}
In order to check the credibility of introducing multi-component recovery model against multi-component injections, we compare the parameter estimation results by applying different recovery models. The injected signals are pure isotropic distribution and mock Galactic-plane distribution. We should note that the Galactic-plane signal also has an isotropic component. The recovery models are a single-component recovery model (Galactic-plane) and a two-component recovery model (Isotropic and Galactic-plane). In injecting and recovering the Galactic-plane distribution, we choose consistent $l_{\rm max}=5$ as the highest l mode. The spherical harmonic components $\mathcal{P}_{lm}$ for the Galactic-plane is given from the mock distribution used in Fig.\@ 1 of Ref.\@ \cite{Tsukada2023}. The exact values of $\bar{\mathcal{P}}_{lm}$ for the mock distribution is given in Tab.\@\ref{tab:mock_galactic}.

\begin{table*}[htbp]
\centering
\setlength{\tabcolsep}{3pt}
\renewcommand{\arraystretch}{1.2}
\scriptsize
\begin{tabular}{c|c|c|c|c|c|c}
$m \backslash \ell$ & 0 & 1 & 2 & 3 & 4 & 5 \\
\hline
-5 &       &       &       &       &       &  0.0078+0.0203$i$ \\
-4 &       &       &       &       &  0.0288-0.0413$i$ & -0.0130-0.0109$i$ \\
-3 &       &       &       & -0.0091+0.0109$i$ & -0.0909+0.0083$i$ & -0.0068+0.0102$i$ \\
-2 &       &       & -0.1613+0.0409$i$ &  0.0238-0.0166$i$ &  0.0396-0.0217$i$ & -0.0156+0.0215$i$ \\
-1 &       &  0.0112-0.0970$i$ &  0.2164-0.0151$i$ & -0.0180+0.0249$i$ &  0.1097-0.0061$i$ & -0.0088+0.0370$i$ \\
 0 &  1.0000 &  0.0218 &  0.0142 & -0.0288 & -0.0895 &  0.0075 \\
 1 &       & -0.0112-0.0970$i$ & -0.2164-0.0151$i$ &  0.0180+0.0249$i$ & -0.1097-0.0061$i$ &  0.0088+0.0370$i$ \\
 2 &       &       & -0.1613-0.0409$i$ &  0.0238+0.0166$i$ &  0.0396+0.0217$i$ & -0.0156-0.0215$i$ \\
 3 &       &       &       &  0.0091+0.0109$i$ &  0.0909+0.0083$i$ &  0.0068+0.0102$i$ \\
 4 &       &       &       &       &  0.0288+0.0413$i$ & -0.0130+0.0109$i$ \\
 5 &       &       &       &       &       & -0.0078+0.0203$i$ \\
\end{tabular}
\caption{$\bar{\mathcal{P}}_{\ell m}$ distribution of mock Galactic-plane distribution used in this paper and \cite{Tsukada2023}. The definition of $\bar{\mathcal{P}}_{\ell m}$ is given by Eq.\@\eqref{eq:Plm_nornalized}}
\label{tab:mock_galactic}
\end{table*}

\subsubsection{$H(f,\vec{\theta})$ model}
\label{subsubsec:Hfmodel}

We assume $H(f,\vec{\theta})$ as a power law model for both injections. The set of model parameters consists of the slope of the spectrum \(\alpha\); \(\vec{\theta} = \{\alpha\}\).
\begin{align}
    H(f, \alpha) = \left(\frac{f}{f_{\mathrm{ref}}}\right)^{\alpha-3},
\end{align}
where $f_{\mathrm{ref}}$ is an arbitrary reference frequency and chosen to be $\SI{25}{Hz}$ in this study. The index parameter $\alpha$ corresponds to the spectral index of $\Omegagw$.

\subsubsection{prior distribution}

The free parameters for both injections (Galactic-plane and Isotropic model) are the amplitude parameter $\epsilon$ and the spectral index $\alpha$ given in Sec.\@ \ref{subsubsec:Hfmodel}. We perform 500 analyses on different noise realizations. The injected parameters are drawn randomly from their prior distributions. We use a log-uniform distribution ranging from $10^{-10}$ to $10^{-5}$ for $\epsilon$. For the index $\alpha$, we use a Gaussian distribution with a mean of $2$ and a standard deviation of $1$. We chose the distribution for $\alpha$ so that it generates values in a range similar to values predicted from potential source models.  For example, $\alpha=0$ corresponds to the \ac{SGWB} from inflationary era \cite{Starobinsky_1979,Turner_1997,Bar_1994} or cosmic strings \cite{Sarangi_2002,Damour_2005,Siemens_2007,Kibble_1976}, and $\alpha=3$ is associated with the \ac{SGWB} from core-collapse supernovae \cite{Buonanno2005,Howell_2004,Sandick_2006,Marassi_2009,Zhu_2010}. Values of $\alpha=2/3, 4$ are associated with the astrophysical \ac{SGWB} originating from \acp{CBC} \cite{Regimbau2008,Rosado2011, Wu2012,Zhu2013,Zhu_2011,Marassi_2011} and magnetars \cite{Regimbau_2001,Marassi_2010,Wu_2013} respectively.  Those injections are made on the simulated noise \ac{CSD} of two LIGO detectors for one year of observation. The spectrum used for simulated noise is the design sensitivity for the Advanced LIGO detectors \cite{Prospects_forALIGO}.

\subsection{Results}

\subsubsection{Posterior distirbution}

\begin{figure}
    \centering
    \includegraphics[width=\linewidth]{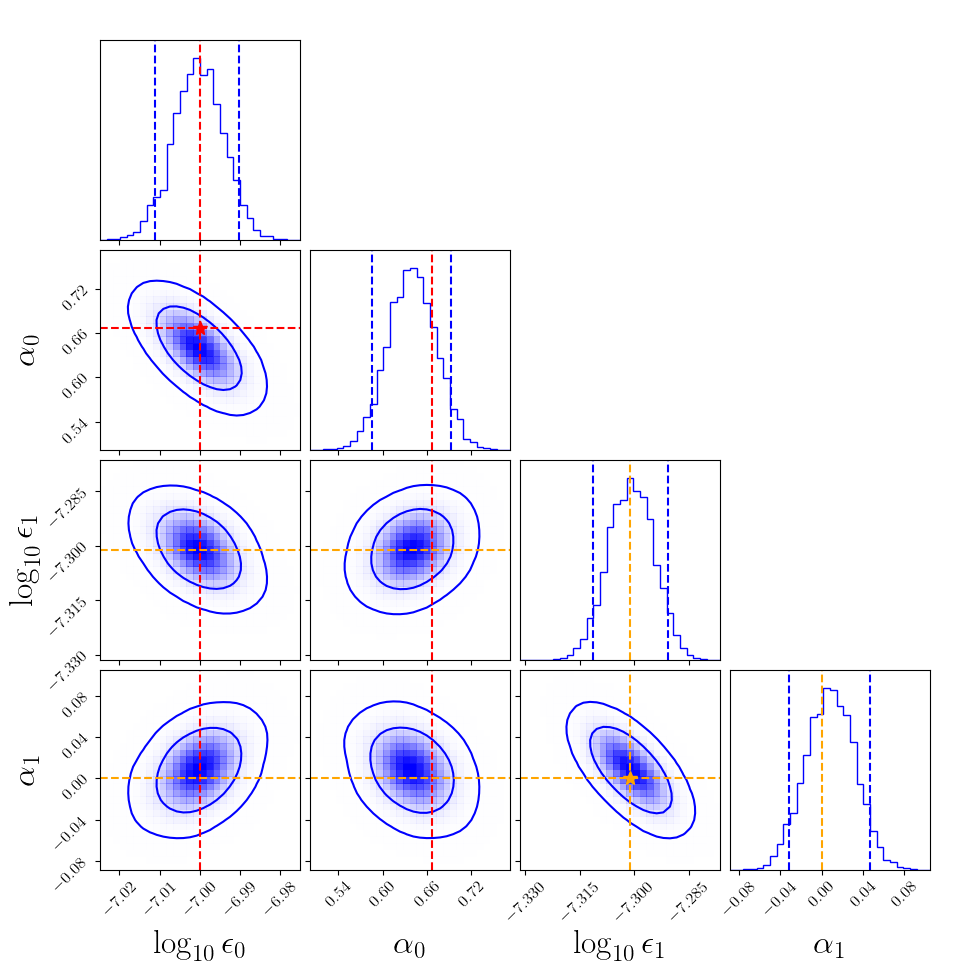}
    \caption{An example posterior distribution of two-component recovery against two-component injection test with $\ln{\mathrm{BF}}\sim2.73\times10^4$. The red marker and lines shows the injection parameters for Galactic-plane injection while yellow ones represents those for Isotropic injection.}
    \label{fig:example_posterior_multi}
\end{figure}

\begin{figure}
    \centering
    \includegraphics[width=.7\linewidth]{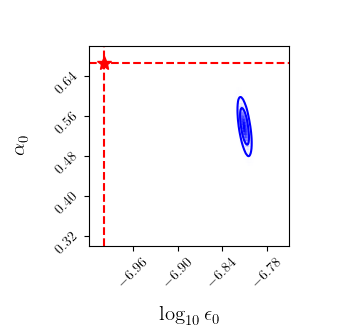}
    \caption{An example posterior distribution of single-component recovery against two-component injection test with $\ln{\mathrm{BF}}\sim1.92\times10^4$. The parameters for Galactic-plane model is recovered only. The red marker and lines shows the injection parameters for Galactic-plane injection. The injection parameter for Isotropic injection is not shown in the figure since it is not the target anisotropy to recover for.}
    \label{fig:example_posterior_single}
\end{figure}

\figref{fig:example_posterior_multi} is an example of the posterior distribution for two-component recovery when the injection is loud enough to have a large Bayes factor for decisive signal evidence. The values of the injected parameters are $(\epsilon_0,\alpha_0)\sim(1.0\times10^{-7},2.0/3.0)$ for Galactic-plane injection indicated by a red star and $(\epsilon_1,\alpha_1)\sim(5.0\times10^{-8},0.0)$ for Isotropic injection indicated by an orange star. The vertical dashed line on the 1-dimensional histograms shows the $(0.05,0.95)$ percentiles for each distribution and the inner circle of the two-dimensional plot shows the $68\%$ contour level while the outer circle is associated with the $95\%$ contour level. As inferred by \figref{fig:example_posterior_multi}, the injected values are precisely recovered within an error of $O(0.1)\%$ for the very loud signal. In contrast, \figref{fig:example_posterior_single}, which is an example of the posterior distribution for single component recovery versus two-component injection, shows a bias in its recovered parameters. The values of the injected parameters are same as the ones in \figref{fig:example_posterior_multi}. The isotropic component of the Galactic-plane signal is recovered with optimal \ac{SNR}$\approx 123$. As the definition of optimal \ac{SNR}, we refer to the Eq.\@9 of Ref.\@ \cite{Callister_2017}. Although we inject a very loud Galactic-plane signal and the logarithmic \ac{BF} ($\sim1.92\times10^4$) for recovery shows a significantly high value, its posterior distribution does not recover the injected signal.

\subsubsection{Probability-probability plot}
\label{sec:pp-plot}

\begin{figure*}
    \includegraphics[width=.5\linewidth]{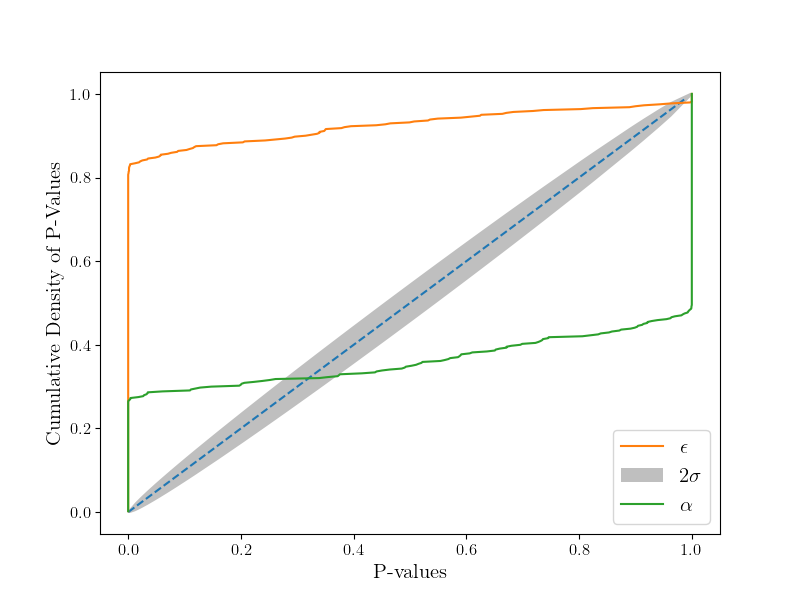}%
    \includegraphics[width=.5\linewidth]{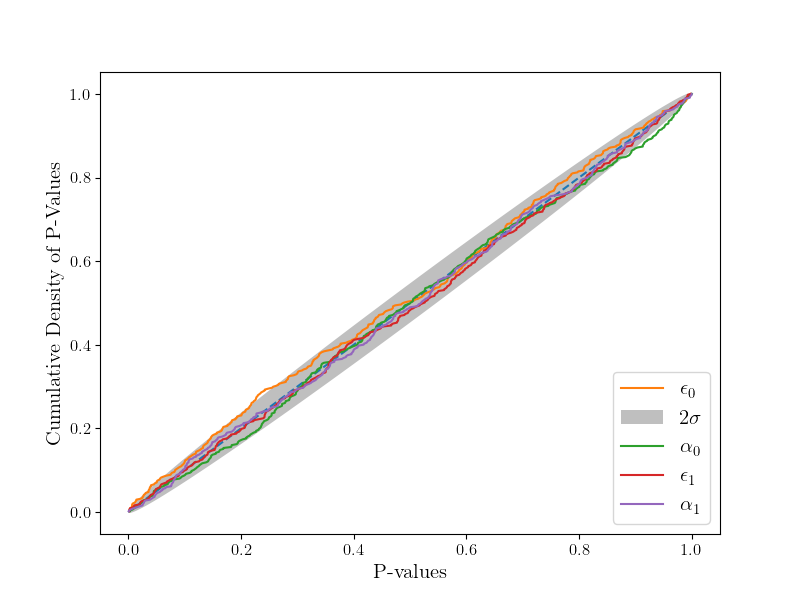}
    \caption{The probability-probability plot for multiple component injection (Isotropic and Galactic-plane). The left plot corresponds to single component (Galactic plane) recovery and the right plot corresponds to multiple component (Isotropic and Galactic-plane) recovery. In the right plot, the index $0$ of $\epsilon$ and $\alpha$ refers to the parameters of Galactic-plane, and index $1$ to that of Isotropic injection. The gray region in both figures is the 2$\sigma$ credible region expected from applying central limit theorem to binomial distribution.}
    \label{fig:pp-plot}
\end{figure*}

We compute the percentile of the value of the injected parameters along the given posterior distribution in each analysis. We then collect 500 different percentiles from each analysis, sort them in ascending order, and compute the cumulative fraction corresponding to each percentile to create the probability-probability plot. \figref{fig:pp-plot} is the probability-probability plots for two different recovery models against injection of multiple components mentioned in Sec.\@\ref{subsubsec:Plmmodel}. For single component recovery (Galactic-plane), we plot the recovery of injected parameters of the same component, which is shown in the left plot of \figref{fig:pp-plot}. We note that the shape of the curve fully depends on the prior assumptions and ranges, and hence the plot on the left of \figref{fig:pp-plot} does not necessarily indicate the level of bias. Nevertheless, that still proves the expected systematic failure in parameter estimation with a single-component recovery model, not an unfortunate realization of the results.

\section{Model Comparison}
\label{sec:BFanalysis}

%\red{I don't understand how the authors were able to assess the impact of "spectral separation" for the model comparison analysis, considering that they fixed the spectral indices for the two different searches (setting the prior distributions
%to be delta functions at the injected spectral index values). By using delta function priors, the spectral index has effectively dropped out of the analysis. (This statement about the impact of spectral separation is also made in Sec. V (Conclusion) refering to Figs.4, 5, 6.)}

In this section, as a further analysis, we examine the changes in \ac{BF} by altering each signal's strength or the noise realizations to demonstrate how much signal strength is required to separate the signals.

\subsection{Fixed Noise Realization}
\begin{figure*}
    \includegraphics[width=.5\linewidth]{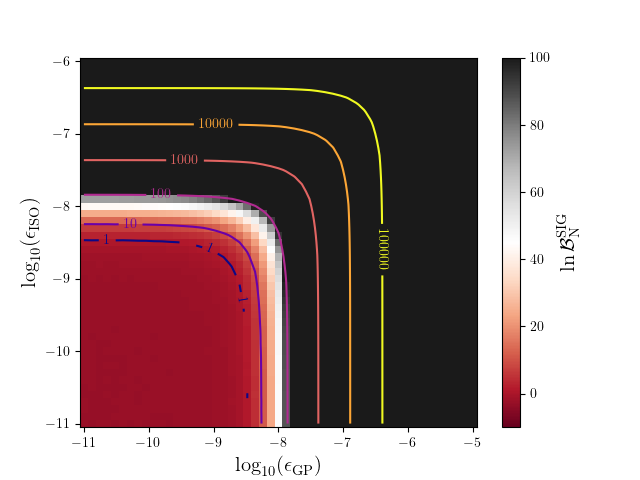}%
    \includegraphics[width=.5\linewidth]{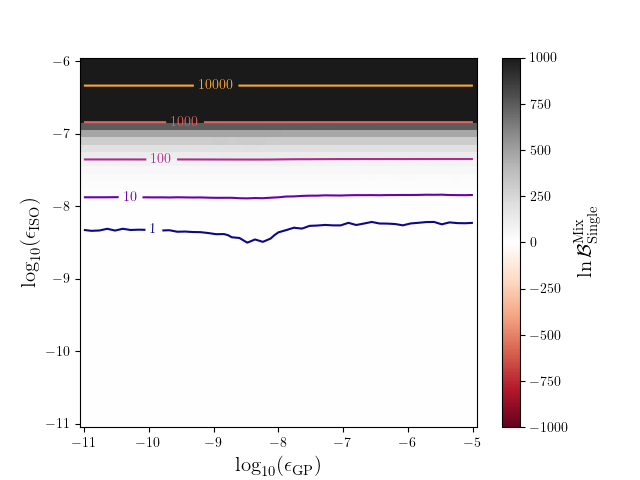}
    \caption{The two dimensional heat maps for natural logarithm of \ac{BF} in the case of injecting different sets of amplitude. Left: \ac{BF}'s heat map of multi-component recovery model against noise model. Right: \ac{BF}'s heat map of multi-component recovery model against single component recovery (Galactic-plane) model.}
    \label{fig:BFmap}
\end{figure*}

Assuming the prior odds for models (let us set $\mathcal{M}_1$ and $\mathcal{M}_2$) are equal, the odds ratio is identical to the \ac{BF} as shown in the following equation
\begin{align}
    O^{\mathcal{M}_1}_{\mathcal{M}_2} = \frac{p(D|\mathcal{M}_1)p(\mathcal{M}_1)}{p(D|\mathcal{M}_2)p(\mathcal{M}_2)} = \frac{p(D|\mathcal{M}_1)}{p(D|\mathcal{M}_2)} = \mathcal{B}^{\mathcal{M}_1}_{\mathcal{M}_2}
\end{align}
where $D$ stands for the data from the detectors. In this section, we describe the significance of a model compared to others by the value of \ac{BF}. According to Ref.\@ \cite{BFcriteria}, evidence against model $\mathcal{M}_2$ is considered very strong when $\ln{\mathcal{B}}^{\mathcal{M}_1}_{\mathcal{M}_2}\sim10$. Also, $\ln{\mathcal{B}}^{\mathcal{M}_1}_{\mathcal{M}_2}\sim1$ is described as "not worth more than a bare mention". To check the variations in \ac{BF} along with the difference in the injected amplitude of the signals, we let the spectral index parameter $\alpha$ be a fixed value of $0.0$ for both injections. In accordance with the injection, a prior distribution of the index parameter is set to be a delta function that peaks at the injected value to observe the effect only from the amplitude of the injections. We apply only one noise realization for this analysis using designed sensitivity for Advanced LIGO detectors \cite{Prospects_forALIGO} and the injected two-component signals (Isotropic and Galactic-plane) with various sets of amplitude parameter $\epsilon$. \figref{fig:BFmap} are two dimensional heat maps of $\ln{\mathcal{B}}^{\mathrm{SIG}}_{\mathrm{N}}$ (left) and $\ln{\mathcal{B}}^{\mathrm{Mix}}_{\mathrm{Single}}$ (right). The left figure is the \ac{BF} heat map of the multi-component recovery model against the noise model. The boundary of $\ln{\mathcal{B}}\sim1$ lies around $\epsilon\sim3\times10^{-9}$ when the amplitude of the other component is weaker by orders of magnitude. When the amplitudes of two components are comparable, the boundary of $\ln{\mathcal{B}}\sim1$ is around $\epsilon\sim2\times10^{-9}$. These $\epsilon$ values correspond to the optimal \ac{SNR} of $2\sim3$. The right figure is the \ac{BF} heat map of the multi-component recovery model against the single component recovery (Galactic-plane) model in the presence of two-component injections. The fluctuation around $\ln{\mathcal{B}}\sim1$ comes from the uncertainty of the noise realization because it acts differently in different noise realizations. 

\subsection{Fixed Injections}
\label{sec:Fixed_injections}
\begin{figure}
    \centering
    \includegraphics[width=\linewidth]{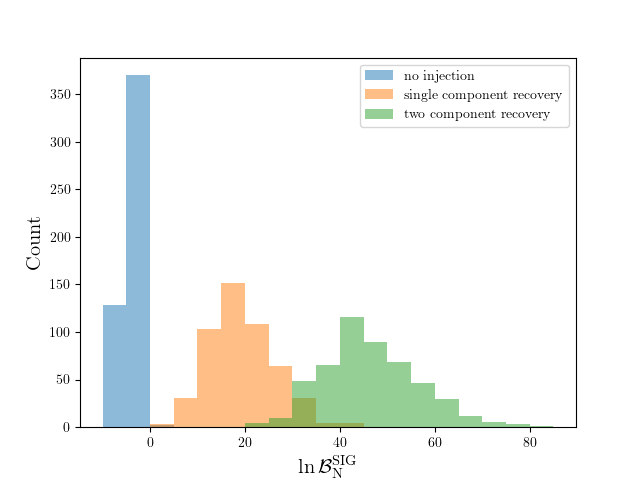}
    \caption{The histogram of $\ln\mathcal{B}_{\rm N}^{\rm SIG}$ for 500 tests with different noise realization each. The blue, orange, and green correspond to two-component recovery with no injection, single-component and two-component recovery for two-component injections.}
    \label{fig:BFhist}
\end{figure}

In this section, we fix the injection amplitude and spectral indices and change the noise realization altering the seed number. The amplitude parameter is chosen so that the isotropic component of the injected distribution has optimal \ac{SNR}$\approx5$. For the injection of Galactic-plane we inject the parameters $(\epsilon_0,\alpha_0)\sim(4.1\times10^{-9},2.0/3.0)$, and $(\epsilon_1,\alpha_1)\sim(4.8\times10^{-9},0.0)$ for the Isotropic injection. We compute the \ac{BF} for 500 different noise realizations with designed sensitivity of Advanced LIGO detectors \cite{Prospects_forALIGO} used in Sec.\@\ref{sec:pp-plot}. We also used the same prior distribution for this analysis, which is a logarithmic uniform distribution ranging from $10^{-10}$ to $10^{-5}$ for $\epsilon$ and a Gaussian distribution with a mean $2$ and a standard deviation of $1$ for $\alpha$. To visualize the distribution of the \acp{BF} for each analysis with several recovery models or injection campaign, we plot the histogram of the \acp{BF}, which is shown in \figref{fig:BFhist}. Compared with the non-injection case and the two-component injection cases, the \acp{BF} for both the single- and two-component recovery model versus the noise model is considerably high. However, \acp{BF} for the two-component recovery model have a higher value than those for single-component recovery in the presence of two-component injections on average, which implies that the former model statistically fits better with the data.

\section{Conclusion}
In this search, we demonstrate injection of multiple-component (isotropic and anisotropic) and its recovery by choosing a different model. We first show the induced bias by choosing the wrong model in \figref{fig:example_posterior_single} while showing the good recovery case by the correct model in \figref{fig:example_posterior_multi}. We show the probability-probability plot in \figref{fig:pp-plot} to demonstrate that choosing the correct model provides statistically consistent results including what is shown in \figref{fig:example_posterior_multi}. It indicates that, in searching for anisotropic \ac{SGWB} with the Bayesian approach presented by \cite{Tsukada2023}, choosing a multiple component recovery model (at least including an isotropic model and a targeted anisotropic model) is necessary since it is highly likely that an isotropic \ac{SGWB} exists in the true sky; otherwise, it causes bias in parameter estimation. It should be noted that the \ac{BF} to the noise model for single component recovery shows significantly high value even though the parameter estimation is clearly biased. We have shown that the two-component recovery model generally provides higher \ac{BF} regardless of noise realizations, which is described in Sec.\@\ref{sec:Fixed_injections}. We also found that this map separation works if \ac{SNR} of one of the signal components is above $2\sim3$ in our simulation setup.

\section{Acknowledgment}
This work has been supported by Japan Society for the Promotion of Science (JSPS) Grants-in-Aid for JSPS Research Fellow (KAKENHI) grant number JP22KJ1033 and Grant-in-Aid for Scientific Research(A) grant number JP18H03698.

The authors are grateful for computational resources provided by the LIGO Laboratory and supported by National Science Foundation Grants PHY-0757058 and PHY-0823459.

L.T acknowledges support from the Nevada Center for Astrophysics.

\bibliography{references}

\begin{thebibliography}{68}%
\makeatletter
\providecommand \@ifxundefined [1]{%
 \@ifx{#1\undefined}
}%
\providecommand \@ifnum [1]{%
 \ifnum #1\expandafter \@firstoftwo
 \else \expandafter \@secondoftwo
 \fi
}%
\providecommand \@ifx [1]{%
 \ifx #1\expandafter \@firstoftwo
 \else \expandafter \@secondoftwo
 \fi
}%
\providecommand \natexlab [1]{#1}%
\providecommand \enquote  [1]{``#1''}%
\providecommand \bibnamefont  [1]{#1}%
\providecommand \bibfnamefont [1]{#1}%
\providecommand \citenamefont [1]{#1}%
\providecommand \href@noop [0]{\@secondoftwo}%
\providecommand \href [0]{\begingroup \@sanitize@url \@href}%
\providecommand \@href[1]{\@@startlink{#1}\@@href}%
\providecommand \@@href[1]{\endgroup#1\@@endlink}%
\providecommand \@sanitize@url [0]{\catcode `\\12\catcode `\$12\catcode `\&12\catcode `\#12\catcode `\^12\catcode `\_12\catcode `\%12\relax}%
\providecommand \@@startlink[1]{}%
\providecommand \@@endlink[0]{}%
\providecommand \url  [0]{\begingroup\@sanitize@url \@url }%
\providecommand \@url [1]{\endgroup\@href {#1}{\urlprefix }}%
\providecommand \urlprefix  [0]{URL }%
\providecommand \Eprint [0]{\href }%
\providecommand \doibase [0]{http://dx.doi.org/}%
\providecommand \selectlanguage [0]{\@gobble}%
\providecommand \bibinfo  [0]{\@secondoftwo}%
\providecommand \bibfield  [0]{\@secondoftwo}%
\providecommand \translation [1]{[#1]}%
\providecommand \BibitemOpen [0]{}%
\providecommand \bibitemStop [0]{}%
\providecommand \bibitemNoStop [0]{.\EOS\space}%
\providecommand \EOS [0]{\spacefactor3000\relax}%
\providecommand \BibitemShut  [1]{\csname bibitem#1\endcsname}%
\let\auto@bib@innerbib\@empty
%</preamble>
\bibitem [{\citenamefont {Regimbau}\ and\ \citenamefont {Mandic}(2008)}]{Regimbau2008}%
  \BibitemOpen
  \bibfield  {author} {\bibinfo {author} {\bibfnamefont {T.}~\bibnamefont {Regimbau}}\ and\ \bibinfo {author} {\bibfnamefont {V.}~\bibnamefont {Mandic}},\ }\href {\doibase 10.1088/0264-9381/25/18/184018} {\bibfield  {journal} {\bibinfo  {journal} {Classical and Quantum Gravity}\ }\textbf {\bibinfo {volume} {25}},\ \bibinfo {pages} {184018} (\bibinfo {year} {2008})}\BibitemShut {NoStop}%
\bibitem [{\citenamefont {Rosado}(2011)}]{Rosado2011}%
  \BibitemOpen
  \bibfield  {author} {\bibinfo {author} {\bibfnamefont {P.~A.}\ \bibnamefont {Rosado}},\ }\href {\doibase 10.1103/PhysRevD.84.084004} {\bibfield  {journal} {\bibinfo  {journal} {Phys. Rev. D}\ }\textbf {\bibinfo {volume} {84}},\ \bibinfo {pages} {084004} (\bibinfo {year} {2011})}\BibitemShut {NoStop}%
\bibitem [{\citenamefont {Wu}\ \emph {et~al.}(2012)\citenamefont {Wu}, \citenamefont {Mandic},\ and\ \citenamefont {Regimbau}}]{Wu2012}%
  \BibitemOpen
  \bibfield  {author} {\bibinfo {author} {\bibfnamefont {C.}~\bibnamefont {Wu}}, \bibinfo {author} {\bibfnamefont {V.}~\bibnamefont {Mandic}}, \ and\ \bibinfo {author} {\bibfnamefont {T.}~\bibnamefont {Regimbau}},\ }\href {\doibase 10.1103/PhysRevD.85.104024} {\bibfield  {journal} {\bibinfo  {journal} {Phys. Rev. D}\ }\textbf {\bibinfo {volume} {85}},\ \bibinfo {pages} {104024} (\bibinfo {year} {2012})}\BibitemShut {NoStop}%
\bibitem [{\citenamefont {Zhu}\ \emph {et~al.}(2013)\citenamefont {Zhu}, \citenamefont {Howell}, \citenamefont {Blair},\ and\ \citenamefont {Zhu}}]{Zhu2013}%
  \BibitemOpen
  \bibfield  {author} {\bibinfo {author} {\bibfnamefont {X.-J.}\ \bibnamefont {Zhu}}, \bibinfo {author} {\bibfnamefont {E.~J.}\ \bibnamefont {Howell}}, \bibinfo {author} {\bibfnamefont {D.~G.}\ \bibnamefont {Blair}}, \ and\ \bibinfo {author} {\bibfnamefont {Z.-H.}\ \bibnamefont {Zhu}},\ }\href {\doibase 10.1093/mnras/stt207} {\bibfield  {journal} {\bibinfo  {journal} {Monthly Notices of the Royal Astronomical Society}\ }\textbf {\bibinfo {volume} {431}},\ \bibinfo {pages} {882} (\bibinfo {year} {2013})},\ \Eprint {http://arxiv.org/abs/https://academic.oup.com/mnras/article-pdf/431/1/882/18243620/stt207.pdf} {https://academic.oup.com/mnras/article-pdf/431/1/882/18243620/stt207.pdf} \BibitemShut {NoStop}%
\bibitem [{\citenamefont {Zhu}\ \emph {et~al.}(2011)\citenamefont {Zhu}, \citenamefont {Howell}, \citenamefont {Regimbau}, \citenamefont {Blair},\ and\ \citenamefont {Zhu}}]{Zhu_2011}%
  \BibitemOpen
  \bibfield  {author} {\bibinfo {author} {\bibfnamefont {X.-J.}\ \bibnamefont {Zhu}}, \bibinfo {author} {\bibfnamefont {E.}~\bibnamefont {Howell}}, \bibinfo {author} {\bibfnamefont {T.}~\bibnamefont {Regimbau}}, \bibinfo {author} {\bibfnamefont {D.}~\bibnamefont {Blair}}, \ and\ \bibinfo {author} {\bibfnamefont {Z.-H.}\ \bibnamefont {Zhu}},\ }\href {\doibase 10.1088/0004-637x/739/2/86} {\bibfield  {journal} {\bibinfo  {journal} {The Astrophysical Journal}\ }\textbf {\bibinfo {volume} {739}},\ \bibinfo {pages} {86} (\bibinfo {year} {2011})}\BibitemShut {NoStop}%
\bibitem [{\citenamefont {Marassi}\ \emph {et~al.}(2011)\citenamefont {Marassi}, \citenamefont {Schneider}, \citenamefont {Corvino}, \citenamefont {Ferrari},\ and\ \citenamefont {Zwart}}]{Marassi_2011}%
  \BibitemOpen
  \bibfield  {author} {\bibinfo {author} {\bibfnamefont {S.}~\bibnamefont {Marassi}}, \bibinfo {author} {\bibfnamefont {R.}~\bibnamefont {Schneider}}, \bibinfo {author} {\bibfnamefont {G.}~\bibnamefont {Corvino}}, \bibinfo {author} {\bibfnamefont {V.}~\bibnamefont {Ferrari}}, \ and\ \bibinfo {author} {\bibfnamefont {S.~P.}\ \bibnamefont {Zwart}},\ }\href {\doibase 10.1103/physrevd.84.124037} {\bibfield  {journal} {\bibinfo  {journal} {Physical Review D}\ }\textbf {\bibinfo {volume} {84}} (\bibinfo {year} {2011}),\ 10.1103/physrevd.84.124037}\BibitemShut {NoStop}%
\bibitem [{\citenamefont {Regimbau}\ and\ \citenamefont {de~Freitas~Pacheco}(2001)}]{Regimbau_2001}%
  \BibitemOpen
  \bibfield  {author} {\bibinfo {author} {\bibfnamefont {T.}~\bibnamefont {Regimbau}}\ and\ \bibinfo {author} {\bibfnamefont {J.~A.}\ \bibnamefont {de~Freitas~Pacheco}},\ }\href {\doibase 10.1051/0004-6361:20011005} {\bibfield  {journal} {\bibinfo  {journal} {Astronomy \& Astrophysics}\ }\textbf {\bibinfo {volume} {376}},\ \bibinfo {pages} {381–385} (\bibinfo {year} {2001})}\BibitemShut {NoStop}%
\bibitem [{\citenamefont {Marassi}\ \emph {et~al.}(2010)\citenamefont {Marassi}, \citenamefont {Ciolfi}, \citenamefont {Schneider}, \citenamefont {Stella},\ and\ \citenamefont {Ferrari}}]{Marassi_2010}%
  \BibitemOpen
  \bibfield  {author} {\bibinfo {author} {\bibfnamefont {S.}~\bibnamefont {Marassi}}, \bibinfo {author} {\bibfnamefont {R.}~\bibnamefont {Ciolfi}}, \bibinfo {author} {\bibfnamefont {R.}~\bibnamefont {Schneider}}, \bibinfo {author} {\bibfnamefont {L.}~\bibnamefont {Stella}}, \ and\ \bibinfo {author} {\bibfnamefont {V.}~\bibnamefont {Ferrari}},\ }\href {\doibase 10.1111/j.1365-2966.2010.17861.x} {\bibfield  {journal} {\bibinfo  {journal} {Monthly Notices of the Royal Astronomical Society}\ }\textbf {\bibinfo {volume} {411}},\ \bibinfo {pages} {2549–2557} (\bibinfo {year} {2010})}\BibitemShut {NoStop}%
\bibitem [{\citenamefont {Wu}\ \emph {et~al.}(2013)\citenamefont {Wu}, \citenamefont {Mandic},\ and\ \citenamefont {Regimbau}}]{Wu_2013}%
  \BibitemOpen
  \bibfield  {author} {\bibinfo {author} {\bibfnamefont {C.-J.}\ \bibnamefont {Wu}}, \bibinfo {author} {\bibfnamefont {V.}~\bibnamefont {Mandic}}, \ and\ \bibinfo {author} {\bibfnamefont {T.}~\bibnamefont {Regimbau}},\ }\href {\doibase 10.1103/PhysRevD.87.042002} {\bibfield  {journal} {\bibinfo  {journal} {Phys. Rev. D}\ }\textbf {\bibinfo {volume} {87}},\ \bibinfo {pages} {042002} (\bibinfo {year} {2013})}\BibitemShut {NoStop}%
\bibitem [{\citenamefont {Buonanno}\ \emph {et~al.}(2005)\citenamefont {Buonanno}, \citenamefont {Sigl}, \citenamefont {Raffelt}, \citenamefont {Janka},\ and\ \citenamefont {Müller}}]{Buonanno2005}%
  \BibitemOpen
  \bibfield  {author} {\bibinfo {author} {\bibfnamefont {A.}~\bibnamefont {Buonanno}}, \bibinfo {author} {\bibfnamefont {G.}~\bibnamefont {Sigl}}, \bibinfo {author} {\bibfnamefont {G.~G.}\ \bibnamefont {Raffelt}}, \bibinfo {author} {\bibfnamefont {H.-T.}\ \bibnamefont {Janka}}, \ and\ \bibinfo {author} {\bibfnamefont {E.}~\bibnamefont {Müller}},\ }\href {\doibase 10.1103/physrevd.72.084001} {\bibfield  {journal} {\bibinfo  {journal} {Physical Review D}\ }\textbf {\bibinfo {volume} {72}} (\bibinfo {year} {2005}),\ 10.1103/physrevd.72.084001}\BibitemShut {NoStop}%
\bibitem [{\citenamefont {Howell}\ \emph {et~al.}(2004)\citenamefont {Howell}, \citenamefont {Coward}, \citenamefont {Burman}, \citenamefont {Blair},\ and\ \citenamefont {Gilmore}}]{Howell_2004}%
  \BibitemOpen
  \bibfield  {author} {\bibinfo {author} {\bibfnamefont {E.}~\bibnamefont {Howell}}, \bibinfo {author} {\bibfnamefont {D.}~\bibnamefont {Coward}}, \bibinfo {author} {\bibfnamefont {R.}~\bibnamefont {Burman}}, \bibinfo {author} {\bibfnamefont {D.}~\bibnamefont {Blair}}, \ and\ \bibinfo {author} {\bibfnamefont {J.}~\bibnamefont {Gilmore}},\ }\href {\doibase 10.1111/j.1365-2966.2004.07863.x} {\bibfield  {journal} {\bibinfo  {journal} {Monthly Notices of the Royal Astronomical Society}\ }\textbf {\bibinfo {volume} {351}},\ \bibinfo {pages} {1237} (\bibinfo {year} {2004})},\ \Eprint {http://arxiv.org/abs/https://academic.oup.com/mnras/article-pdf/351/4/1237/3924132/351-4-1237.pdf} {https://academic.oup.com/mnras/article-pdf/351/4/1237/3924132/351-4-1237.pdf} \BibitemShut {NoStop}%
\bibitem [{\citenamefont {Sandick}\ \emph {et~al.}(2006)\citenamefont {Sandick}, \citenamefont {Olive}, \citenamefont {Daigne},\ and\ \citenamefont {Vangioni}}]{Sandick_2006}%
  \BibitemOpen
  \bibfield  {author} {\bibinfo {author} {\bibfnamefont {P.}~\bibnamefont {Sandick}}, \bibinfo {author} {\bibfnamefont {K.~A.}\ \bibnamefont {Olive}}, \bibinfo {author} {\bibfnamefont {F.}~\bibnamefont {Daigne}}, \ and\ \bibinfo {author} {\bibfnamefont {E.}~\bibnamefont {Vangioni}},\ }\href {\doibase 10.1103/physrevd.73.104024} {\bibfield  {journal} {\bibinfo  {journal} {Physical Review D}\ }\textbf {\bibinfo {volume} {73}} (\bibinfo {year} {2006}),\ 10.1103/physrevd.73.104024}\BibitemShut {NoStop}%
\bibitem [{\citenamefont {Marassi}\ \emph {et~al.}(2009)\citenamefont {Marassi}, \citenamefont {Schneider},\ and\ \citenamefont {Ferrari}}]{Marassi_2009}%
  \BibitemOpen
  \bibfield  {author} {\bibinfo {author} {\bibfnamefont {S.}~\bibnamefont {Marassi}}, \bibinfo {author} {\bibfnamefont {R.}~\bibnamefont {Schneider}}, \ and\ \bibinfo {author} {\bibfnamefont {V.}~\bibnamefont {Ferrari}},\ }\href {\doibase 10.1111/j.1365-2966.2009.15120.x} {\bibfield  {journal} {\bibinfo  {journal} {Monthly Notices of the Royal Astronomical Society}\ }\textbf {\bibinfo {volume} {398}},\ \bibinfo {pages} {293–302} (\bibinfo {year} {2009})}\BibitemShut {NoStop}%
\bibitem [{\citenamefont {Zhu}\ \emph {et~al.}(2010)\citenamefont {Zhu}, \citenamefont {Howell},\ and\ \citenamefont {Blair}}]{Zhu_2010}%
  \BibitemOpen
  \bibfield  {author} {\bibinfo {author} {\bibfnamefont {X.-J.}\ \bibnamefont {Zhu}}, \bibinfo {author} {\bibfnamefont {E.}~\bibnamefont {Howell}}, \ and\ \bibinfo {author} {\bibfnamefont {D.}~\bibnamefont {Blair}},\ }\href {\doibase 10.1111/j.1745-3933.2010.00965.x} {\bibfield  {journal} {\bibinfo  {journal} {Monthly Notices of the Royal Astronomical Society: Letters}\ }\textbf {\bibinfo {volume} {409}},\ \bibinfo {pages} {L132–L136} (\bibinfo {year} {2010})}\BibitemShut {NoStop}%
\bibitem [{\citenamefont {Sarangi}\ and\ \citenamefont {Tye}(2002)}]{Sarangi_2002}%
  \BibitemOpen
  \bibfield  {author} {\bibinfo {author} {\bibfnamefont {S.}~\bibnamefont {Sarangi}}\ and\ \bibinfo {author} {\bibfnamefont {S.-H.}\ \bibnamefont {Tye}},\ }\href {\doibase 10.1016/s0370-2693(02)01824-5} {\bibfield  {journal} {\bibinfo  {journal} {Physics Letters B}\ }\textbf {\bibinfo {volume} {536}},\ \bibinfo {pages} {185–192} (\bibinfo {year} {2002})}\BibitemShut {NoStop}%
\bibitem [{\citenamefont {Damour}\ and\ \citenamefont {Vilenkin}(2005)}]{Damour_2005}%
  \BibitemOpen
  \bibfield  {author} {\bibinfo {author} {\bibfnamefont {T.}~\bibnamefont {Damour}}\ and\ \bibinfo {author} {\bibfnamefont {A.}~\bibnamefont {Vilenkin}},\ }\href {\doibase 10.1103/PhysRevD.71.063510} {\bibfield  {journal} {\bibinfo  {journal} {Phys. Rev. D}\ }\textbf {\bibinfo {volume} {71}},\ \bibinfo {pages} {063510} (\bibinfo {year} {2005})}\BibitemShut {NoStop}%
\bibitem [{\citenamefont {Siemens}\ \emph {et~al.}(2007)\citenamefont {Siemens}, \citenamefont {Mandic},\ and\ \citenamefont {Creighton}}]{Siemens_2007}%
  \BibitemOpen
  \bibfield  {author} {\bibinfo {author} {\bibfnamefont {X.}~\bibnamefont {Siemens}}, \bibinfo {author} {\bibfnamefont {V.}~\bibnamefont {Mandic}}, \ and\ \bibinfo {author} {\bibfnamefont {J.}~\bibnamefont {Creighton}},\ }\href {\doibase 10.1103/physrevlett.98.111101} {\bibfield  {journal} {\bibinfo  {journal} {Physical Review Letters}\ }\textbf {\bibinfo {volume} {98}} (\bibinfo {year} {2007}),\ 10.1103/physrevlett.98.111101}\BibitemShut {NoStop}%
\bibitem [{\citenamefont {Kibble}(1976)}]{Kibble_1976}%
  \BibitemOpen
  \bibfield  {author} {\bibinfo {author} {\bibfnamefont {T.~W.~B.}\ \bibnamefont {Kibble}},\ }\href {\doibase 10.1088/0305-4470/9/8/029} {\bibfield  {journal} {\bibinfo  {journal} {Journal of Physics A: Mathematical and General}\ }\textbf {\bibinfo {volume} {9}},\ \bibinfo {pages} {1387} (\bibinfo {year} {1976})}\BibitemShut {NoStop}%
\bibitem [{\citenamefont {Starobinsky}(1979)}]{Starobinsky_1979}%
  \BibitemOpen
  \bibfield  {author} {\bibinfo {author} {\bibfnamefont {A.}~\bibnamefont {Starobinsky}},\ }\href@noop {} {\bibfield  {journal} {\bibinfo  {journal} {JETP Lett.}\ }\textbf {\bibinfo {volume} {30}},\ \bibinfo {pages} {682} (\bibinfo {year} {1979})}\BibitemShut {NoStop}%
\bibitem [{\citenamefont {Turner}(1997)}]{Turner_1997}%
  \BibitemOpen
  \bibfield  {author} {\bibinfo {author} {\bibfnamefont {M.~S.}\ \bibnamefont {Turner}},\ }\href {\doibase 10.1103/physrevd.55.r435} {\bibfield  {journal} {\bibinfo  {journal} {Physical Review D}\ }\textbf {\bibinfo {volume} {55}},\ \bibinfo {pages} {R435–R439} (\bibinfo {year} {1997})}\BibitemShut {NoStop}%
\bibitem [{\citenamefont {Bar-Kana}(1994)}]{Bar_1994}%
  \BibitemOpen
  \bibfield  {author} {\bibinfo {author} {\bibfnamefont {R.}~\bibnamefont {Bar-Kana}},\ }\href {\doibase 10.1103/physrevd.50.1157} {\bibfield  {journal} {\bibinfo  {journal} {Physical Review D}\ }\textbf {\bibinfo {volume} {50}},\ \bibinfo {pages} {1157–1160} (\bibinfo {year} {1994})}\BibitemShut {NoStop}%
\bibitem [{\citenamefont {Sasaki}\ \emph {et~al.}(2016)\citenamefont {Sasaki}, \citenamefont {Suyama}, \citenamefont {Tanaka},\ and\ \citenamefont {Yokoyama}}]{Sasaki_2016}%
  \BibitemOpen
  \bibfield  {author} {\bibinfo {author} {\bibfnamefont {M.}~\bibnamefont {Sasaki}}, \bibinfo {author} {\bibfnamefont {T.}~\bibnamefont {Suyama}}, \bibinfo {author} {\bibfnamefont {T.}~\bibnamefont {Tanaka}}, \ and\ \bibinfo {author} {\bibfnamefont {S.}~\bibnamefont {Yokoyama}},\ }\href {\doibase 10.1103/PhysRevLett.117.061101} {\bibfield  {journal} {\bibinfo  {journal} {Phys. Rev. Lett.}\ }\textbf {\bibinfo {volume} {117}},\ \bibinfo {pages} {061101} (\bibinfo {year} {2016})}\BibitemShut {NoStop}%
\bibitem [{\citenamefont {Mandic}\ \emph {et~al.}(2016)\citenamefont {Mandic}, \citenamefont {Bird},\ and\ \citenamefont {Cholis}}]{Mandic_2016}%
  \BibitemOpen
  \bibfield  {author} {\bibinfo {author} {\bibfnamefont {V.}~\bibnamefont {Mandic}}, \bibinfo {author} {\bibfnamefont {S.}~\bibnamefont {Bird}}, \ and\ \bibinfo {author} {\bibfnamefont {I.}~\bibnamefont {Cholis}},\ }\href {\doibase 10.1103/physrevlett.117.201102} {\bibfield  {journal} {\bibinfo  {journal} {Physical Review Letters}\ }\textbf {\bibinfo {volume} {117}} (\bibinfo {year} {2016}),\ 10.1103/physrevlett.117.201102}\BibitemShut {NoStop}%
\bibitem [{\citenamefont {Wang}\ \emph {et~al.}(2018)\citenamefont {Wang}, \citenamefont {Wang}, \citenamefont {Huang},\ and\ \citenamefont {Li}}]{Wang_2018}%
  \BibitemOpen
  \bibfield  {author} {\bibinfo {author} {\bibfnamefont {S.}~\bibnamefont {Wang}}, \bibinfo {author} {\bibfnamefont {Y.-F.}\ \bibnamefont {Wang}}, \bibinfo {author} {\bibfnamefont {Q.-G.}\ \bibnamefont {Huang}}, \ and\ \bibinfo {author} {\bibfnamefont {T.~G.}\ \bibnamefont {Li}},\ }\href {\doibase 10.1103/physrevlett.120.191102} {\bibfield  {journal} {\bibinfo  {journal} {Physical Review Letters}\ }\textbf {\bibinfo {volume} {120}} (\bibinfo {year} {2018}),\ 10.1103/physrevlett.120.191102}\BibitemShut {NoStop}%
\bibitem [{\citenamefont {Aasi}\ \emph {et~al.}(2015)\citenamefont {Aasi} \emph {et~al.}}]{DetectorALIGO}%
  \BibitemOpen
  \bibfield  {author} {\bibinfo {author} {\bibfnamefont {J.}~\bibnamefont {Aasi}} \emph {et~al.} (\bibinfo {collaboration} {LIGO Scientific}),\ }\href {\doibase 10.1088/0264-9381/32/7/074001} {\bibfield  {journal} {\bibinfo  {journal} {Class. Quant. Grav.}\ }\textbf {\bibinfo {volume} {32}},\ \bibinfo {pages} {074001} (\bibinfo {year} {2015})},\ \Eprint {http://arxiv.org/abs/1411.4547} {arXiv:1411.4547 [gr-qc]} \BibitemShut {NoStop}%
\bibitem [{\citenamefont {Acernese}\ \emph {et~al.}(2015)\citenamefont {Acernese} \emph {et~al.}}]{DetectorAVirgo}%
  \BibitemOpen
  \bibfield  {author} {\bibinfo {author} {\bibfnamefont {F.}~\bibnamefont {Acernese}} \emph {et~al.} (\bibinfo {collaboration} {VIRGO}),\ }\href {\doibase 10.1088/0264-9381/32/2/024001} {\bibfield  {journal} {\bibinfo  {journal} {Class. Quant. Grav.}\ }\textbf {\bibinfo {volume} {32}},\ \bibinfo {pages} {024001} (\bibinfo {year} {2015})},\ \Eprint {http://arxiv.org/abs/1408.3978} {arXiv:1408.3978 [gr-qc]} \BibitemShut {NoStop}%
\bibitem [{\citenamefont {Akutsu}\ \emph {et~al.}(2020)\citenamefont {Akutsu} \emph {et~al.}}]{DetectorKAGRA}%
  \BibitemOpen
  \bibfield  {author} {\bibinfo {author} {\bibfnamefont {T.}~\bibnamefont {Akutsu}} \emph {et~al.},\ }\href {\doibase 10.1093/ptep/ptaa125} {\bibfield  {journal} {\bibinfo  {journal} {Progress of Theoretical and Experimental Physics}\ }\textbf {\bibinfo {volume} {2021}},\ \bibinfo {pages} {05A101} (\bibinfo {year} {2020})},\ \Eprint {http://arxiv.org/abs/https://academic.oup.com/ptep/article-pdf/2021/5/05A101/37974994/ptaa125.pdf} {https://academic.oup.com/ptep/article-pdf/2021/5/05A101/37974994/ptaa125.pdf} \BibitemShut {NoStop}%
\bibitem [{\citenamefont {Abbott}\ \emph {et~al.}(2021{\natexlab{a}})\citenamefont {Abbott} \emph {et~al.}}]{O3Isotropic}%
  \BibitemOpen
  \bibfield  {author} {\bibinfo {author} {\bibfnamefont {R.}~\bibnamefont {Abbott}} \emph {et~al.},\ }\href {\doibase 10.1103/physrevd.104.022004} {\bibfield  {journal} {\bibinfo  {journal} {Physical Review D}\ }\textbf {\bibinfo {volume} {104}} (\bibinfo {year} {2021}{\natexlab{a}}),\ 10.1103/physrevd.104.022004}\BibitemShut {NoStop}%
\bibitem [{\citenamefont {Xu}\ \emph {et~al.}(2023)\citenamefont {Xu} \emph {et~al.}}]{CPTA2023Jun}%
  \BibitemOpen
  \bibfield  {author} {\bibinfo {author} {\bibfnamefont {H.}~\bibnamefont {Xu}} \emph {et~al.},\ }\href {\doibase 10.1088/1674-4527/acdfa5} {\bibfield  {journal} {\bibinfo  {journal} {Research in Astronomy and Astrophysics}\ }\textbf {\bibinfo {volume} {23}},\ \bibinfo {pages} {075024} (\bibinfo {year} {2023})}\BibitemShut {NoStop}%
\bibitem [{\citenamefont {Agazie}\ \emph {et~al.}(2023)\citenamefont {Agazie} \emph {et~al.}}]{nanoGrav2023Jun}%
  \BibitemOpen
  \bibfield  {author} {\bibinfo {author} {\bibfnamefont {G.}~\bibnamefont {Agazie}} \emph {et~al.},\ }\href {\doibase 10.3847/2041-8213/acdac6} {\bibfield  {journal} {\bibinfo  {journal} {The Astrophysical Journal Letters}\ }\textbf {\bibinfo {volume} {951}},\ \bibinfo {pages} {L8} (\bibinfo {year} {2023})}\BibitemShut {NoStop}%
\bibitem [{\citenamefont {Antoniadis}\ \emph {et~al.}(2023)\citenamefont {Antoniadis} \emph {et~al.}}]{EPTA2023Jun}%
  \BibitemOpen
  \bibfield  {author} {\bibinfo {author} {\bibfnamefont {J.}~\bibnamefont {Antoniadis}} \emph {et~al.},\ }\href {\doibase 10.1051/0004-6361/202346844} {\bibfield  {journal} {\bibinfo  {journal} {Astronomy and Astrophysics}\ }\textbf {\bibinfo {volume} {678}},\ \bibinfo {pages} {A50} (\bibinfo {year} {2023})}\BibitemShut {NoStop}%
\bibitem [{\citenamefont {Reardon}\ \emph {et~al.}(2023)\citenamefont {Reardon} \emph {et~al.}}]{PPTA2023Jun}%
  \BibitemOpen
  \bibfield  {author} {\bibinfo {author} {\bibfnamefont {D.~J.}\ \bibnamefont {Reardon}} \emph {et~al.},\ }\href {\doibase 10.3847/2041-8213/acdd02} {\bibfield  {journal} {\bibinfo  {journal} {The Astrophysical Journal Letters}\ }\textbf {\bibinfo {volume} {951}},\ \bibinfo {pages} {L6} (\bibinfo {year} {2023})}\BibitemShut {NoStop}%
\bibitem [{\citenamefont {Contaldi}(2017)}]{Contaldi_2017}%
  \BibitemOpen
  \bibfield  {author} {\bibinfo {author} {\bibfnamefont {C.~R.}\ \bibnamefont {Contaldi}},\ }\href {\doibase 10.1016/j.physletb.2017.05.020} {\bibfield  {journal} {\bibinfo  {journal} {Physics Letters B}\ }\textbf {\bibinfo {volume} {771}},\ \bibinfo {pages} {9–12} (\bibinfo {year} {2017})}\BibitemShut {NoStop}%
\bibitem [{\citenamefont {Jenkins}\ \emph {et~al.}(2019)\citenamefont {Jenkins}, \citenamefont {Romano},\ and\ \citenamefont {Sakellariadou}}]{Jenkins_2019_oct}%
  \BibitemOpen
  \bibfield  {author} {\bibinfo {author} {\bibfnamefont {A.~C.}\ \bibnamefont {Jenkins}}, \bibinfo {author} {\bibfnamefont {J.~D.}\ \bibnamefont {Romano}}, \ and\ \bibinfo {author} {\bibfnamefont {M.}~\bibnamefont {Sakellariadou}},\ }\href {\doibase 10.1103/physrevd.100.083501} {\bibfield  {journal} {\bibinfo  {journal} {Physical Review D}\ }\textbf {\bibinfo {volume} {100}} (\bibinfo {year} {2019}),\ 10.1103/physrevd.100.083501}\BibitemShut {NoStop}%
\bibitem [{\citenamefont {Jenkins}\ and\ \citenamefont {Sakellariadou}(2019)}]{Jenkins_2019_Sep}%
  \BibitemOpen
  \bibfield  {author} {\bibinfo {author} {\bibfnamefont {A.~C.}\ \bibnamefont {Jenkins}}\ and\ \bibinfo {author} {\bibfnamefont {M.}~\bibnamefont {Sakellariadou}},\ }\href {\doibase 10.1103/PhysRevD.100.063508} {\bibfield  {journal} {\bibinfo  {journal} {Phys. Rev. D}\ }\textbf {\bibinfo {volume} {100}},\ \bibinfo {pages} {063508} (\bibinfo {year} {2019})}\BibitemShut {NoStop}%
\bibitem [{\citenamefont {Bertacca}\ \emph {et~al.}(2020)\citenamefont {Bertacca}, \citenamefont {Ricciardone}, \citenamefont {Bellomo}, \citenamefont {Jenkins}, \citenamefont {Matarrese}, \citenamefont {Raccanelli}, \citenamefont {Regimbau},\ and\ \citenamefont {Sakellariadou}}]{Bertacca_2020}%
  \BibitemOpen
  \bibfield  {author} {\bibinfo {author} {\bibfnamefont {D.}~\bibnamefont {Bertacca}}, \bibinfo {author} {\bibfnamefont {A.}~\bibnamefont {Ricciardone}}, \bibinfo {author} {\bibfnamefont {N.}~\bibnamefont {Bellomo}}, \bibinfo {author} {\bibfnamefont {A.~C.}\ \bibnamefont {Jenkins}}, \bibinfo {author} {\bibfnamefont {S.}~\bibnamefont {Matarrese}}, \bibinfo {author} {\bibfnamefont {A.}~\bibnamefont {Raccanelli}}, \bibinfo {author} {\bibfnamefont {T.}~\bibnamefont {Regimbau}}, \ and\ \bibinfo {author} {\bibfnamefont {M.}~\bibnamefont {Sakellariadou}},\ }\href {\doibase 10.1103/physrevd.101.103513} {\bibfield  {journal} {\bibinfo  {journal} {Physical Review D}\ }\textbf {\bibinfo {volume} {101}} (\bibinfo {year} {2020}),\ 10.1103/physrevd.101.103513}\BibitemShut {NoStop}%
\bibitem [{\citenamefont {Cusin}\ \emph {et~al.}(2017)\citenamefont {Cusin}, \citenamefont {Pitrou},\ and\ \citenamefont {Uzan}}]{Cusin_2017}%
  \BibitemOpen
  \bibfield  {author} {\bibinfo {author} {\bibfnamefont {G.}~\bibnamefont {Cusin}}, \bibinfo {author} {\bibfnamefont {C.}~\bibnamefont {Pitrou}}, \ and\ \bibinfo {author} {\bibfnamefont {J.-P.}\ \bibnamefont {Uzan}},\ }\href {\doibase 10.1103/physrevd.96.103019} {\bibfield  {journal} {\bibinfo  {journal} {Physical Review D}\ }\textbf {\bibinfo {volume} {96}} (\bibinfo {year} {2017}),\ 10.1103/physrevd.96.103019}\BibitemShut {NoStop}%
\bibitem [{\citenamefont {Cusin}\ \emph {et~al.}(2018{\natexlab{a}})\citenamefont {Cusin}, \citenamefont {Pitrou},\ and\ \citenamefont {Uzan}}]{Cusin_2018}%
  \BibitemOpen
  \bibfield  {author} {\bibinfo {author} {\bibfnamefont {G.}~\bibnamefont {Cusin}}, \bibinfo {author} {\bibfnamefont {C.}~\bibnamefont {Pitrou}}, \ and\ \bibinfo {author} {\bibfnamefont {J.-P.}\ \bibnamefont {Uzan}},\ }\href {\doibase 10.1103/physrevd.97.123527} {\bibfield  {journal} {\bibinfo  {journal} {Physical Review D}\ }\textbf {\bibinfo {volume} {97}} (\bibinfo {year} {2018}{\natexlab{a}}),\ 10.1103/physrevd.97.123527}\BibitemShut {NoStop}%
\bibitem [{\citenamefont {Cusin}\ \emph {et~al.}(2018{\natexlab{b}})\citenamefont {Cusin}, \citenamefont {Dvorkin}, \citenamefont {Pitrou},\ and\ \citenamefont {Uzan}}]{Cusin_2018_2}%
  \BibitemOpen
  \bibfield  {author} {\bibinfo {author} {\bibfnamefont {G.}~\bibnamefont {Cusin}}, \bibinfo {author} {\bibfnamefont {I.}~\bibnamefont {Dvorkin}}, \bibinfo {author} {\bibfnamefont {C.}~\bibnamefont {Pitrou}}, \ and\ \bibinfo {author} {\bibfnamefont {J.-P.}\ \bibnamefont {Uzan}},\ }\href {\doibase 10.1103/physrevlett.120.231101} {\bibfield  {journal} {\bibinfo  {journal} {Physical Review Letters}\ }\textbf {\bibinfo {volume} {120}} (\bibinfo {year} {2018}{\natexlab{b}}),\ 10.1103/physrevlett.120.231101}\BibitemShut {NoStop}%
\bibitem [{\citenamefont {Cusin}\ \emph {et~al.}(2019)\citenamefont {Cusin}, \citenamefont {Dvorkin}, \citenamefont {Pitrou},\ and\ \citenamefont {Uzan}}]{Cusin_2019}%
  \BibitemOpen
  \bibfield  {author} {\bibinfo {author} {\bibfnamefont {G.}~\bibnamefont {Cusin}}, \bibinfo {author} {\bibfnamefont {I.}~\bibnamefont {Dvorkin}}, \bibinfo {author} {\bibfnamefont {C.}~\bibnamefont {Pitrou}}, \ and\ \bibinfo {author} {\bibfnamefont {J.-P.}\ \bibnamefont {Uzan}},\ }\href {\doibase 10.1103/physrevd.100.063004} {\bibfield  {journal} {\bibinfo  {journal} {Physical Review D}\ }\textbf {\bibinfo {volume} {100}} (\bibinfo {year} {2019}),\ 10.1103/physrevd.100.063004}\BibitemShut {NoStop}%
\bibitem [{\citenamefont {Pitrou}\ \emph {et~al.}(2020)\citenamefont {Pitrou}, \citenamefont {Cusin},\ and\ \citenamefont {Uzan}}]{Pitrou_2020}%
  \BibitemOpen
  \bibfield  {author} {\bibinfo {author} {\bibfnamefont {C.}~\bibnamefont {Pitrou}}, \bibinfo {author} {\bibfnamefont {G.}~\bibnamefont {Cusin}}, \ and\ \bibinfo {author} {\bibfnamefont {J.-P.}\ \bibnamefont {Uzan}},\ }\href {\doibase 10.1103/physrevd.101.081301} {\bibfield  {journal} {\bibinfo  {journal} {Physical Review D}\ }\textbf {\bibinfo {volume} {101}} (\bibinfo {year} {2020}),\ 10.1103/physrevd.101.081301}\BibitemShut {NoStop}%
\bibitem [{\citenamefont {Cañas-Herrera}\ \emph {et~al.}(2020)\citenamefont {Cañas-Herrera}, \citenamefont {Contigiani},\ and\ \citenamefont {Vardanyan}}]{Ca_as_Herrera_2020}%
  \BibitemOpen
  \bibfield  {author} {\bibinfo {author} {\bibfnamefont {G.}~\bibnamefont {Cañas-Herrera}}, \bibinfo {author} {\bibfnamefont {O.}~\bibnamefont {Contigiani}}, \ and\ \bibinfo {author} {\bibfnamefont {V.}~\bibnamefont {Vardanyan}},\ }\href {\doibase 10.1103/physrevd.102.043513} {\bibfield  {journal} {\bibinfo  {journal} {Physical Review D}\ }\textbf {\bibinfo {volume} {102}} (\bibinfo {year} {2020}),\ 10.1103/physrevd.102.043513}\BibitemShut {NoStop}%
\bibitem [{\citenamefont {Geller}\ \emph {et~al.}(2018)\citenamefont {Geller}, \citenamefont {Hook}, \citenamefont {Sundrum},\ and\ \citenamefont {Tsai}}]{Geller_2018}%
  \BibitemOpen
  \bibfield  {author} {\bibinfo {author} {\bibfnamefont {M.}~\bibnamefont {Geller}}, \bibinfo {author} {\bibfnamefont {A.}~\bibnamefont {Hook}}, \bibinfo {author} {\bibfnamefont {R.}~\bibnamefont {Sundrum}}, \ and\ \bibinfo {author} {\bibfnamefont {Y.}~\bibnamefont {Tsai}},\ }\href {\doibase 10.1103/physrevlett.121.201303} {\bibfield  {journal} {\bibinfo  {journal} {Physical Review Letters}\ }\textbf {\bibinfo {volume} {121}} (\bibinfo {year} {2018}),\ 10.1103/physrevlett.121.201303}\BibitemShut {NoStop}%
\bibitem [{\citenamefont {Abbott}\ \emph {et~al.}(2017{\natexlab{a}})\citenamefont {Abbott} \emph {et~al.}}]{Abbott_2017}%
  \BibitemOpen
  \bibfield  {author} {\bibinfo {author} {\bibfnamefont {B.~P.}\ \bibnamefont {Abbott}} \emph {et~al.},\ }\href {\doibase 10.3847/1538-4357/aa86f0} {\bibfield  {journal} {\bibinfo  {journal} {The Astrophysical Journal}\ }\textbf {\bibinfo {volume} {847}},\ \bibinfo {pages} {47} (\bibinfo {year} {2017}{\natexlab{a}})}\BibitemShut {NoStop}%
\bibitem [{\citenamefont {Abbott}\ \emph {et~al.}(2017{\natexlab{b}})\citenamefont {Abbott} \emph {et~al.}}]{Abbott_2017_Jun}%
  \BibitemOpen
  \bibfield  {author} {\bibinfo {author} {\bibfnamefont {B.~P.}\ \bibnamefont {Abbott}} \emph {et~al.} (\bibinfo {collaboration} {LIGO Scientific Collaboration and Virgo Collaboration}),\ }\href {\doibase 10.1103/PhysRevD.95.122003} {\bibfield  {journal} {\bibinfo  {journal} {Phys. Rev. D}\ }\textbf {\bibinfo {volume} {95}},\ \bibinfo {pages} {122003} (\bibinfo {year} {2017}{\natexlab{b}})}\BibitemShut {NoStop}%
\bibitem [{\citenamefont {Abbott}\ \emph {et~al.}(2021{\natexlab{b}})\citenamefont {Abbott} \emph {et~al.}}]{O3Anisotropic}%
  \BibitemOpen
  \bibfield  {author} {\bibinfo {author} {\bibfnamefont {R.}~\bibnamefont {Abbott}} \emph {et~al.} (\bibinfo {collaboration} {LIGO Scientific Collaboration, Virgo Collaboration, and KAGRA Collaboration}),\ }\href {\doibase 10.1103/PhysRevD.104.022005} {\bibfield  {journal} {\bibinfo  {journal} {Phys. Rev. D}\ }\textbf {\bibinfo {volume} {104}},\ \bibinfo {pages} {022005} (\bibinfo {year} {2021}{\natexlab{b}})}\BibitemShut {NoStop}%
\bibitem [{\citenamefont {Abbott}\ \emph {et~al.}(2007)\citenamefont {Abbott} \emph {et~al.}}]{BBR_LIGO}%
  \BibitemOpen
  \bibfield  {author} {\bibinfo {author} {\bibfnamefont {B.}~\bibnamefont {Abbott}} \emph {et~al.} (\bibinfo {collaboration} {LIGO Scientific Collaboration}),\ }\href {\doibase 10.1103/PhysRevD.76.082003} {\bibfield  {journal} {\bibinfo  {journal} {Phys. Rev. D}\ }\textbf {\bibinfo {volume} {76}},\ \bibinfo {pages} {082003} (\bibinfo {year} {2007})}\BibitemShut {NoStop}%
\bibitem [{\citenamefont {Ballmer}(2006{\natexlab{a}})}]{BBR_Ballmer}%
  \BibitemOpen
  \bibfield  {author} {\bibinfo {author} {\bibfnamefont {S.~W.}\ \bibnamefont {Ballmer}},\ }\href {\doibase 10.1088/0264-9381/23/8/S23} {\bibfield  {journal} {\bibinfo  {journal} {Classical and Quantum Gravity}\ }\textbf {\bibinfo {volume} {23}},\ \bibinfo {pages} {S179} (\bibinfo {year} {2006}{\natexlab{a}})}\BibitemShut {NoStop}%
\bibitem [{\citenamefont {Ballmer}(2006{\natexlab{b}})}]{NBR_Ballmer}%
  \BibitemOpen
  \bibfield  {author} {\bibinfo {author} {\bibfnamefont {S.~W.}\ \bibnamefont {Ballmer}},\ }\emph {\bibinfo {title} {LIGO interferometer operating at design sensitivity with application to gravitational radiometry}},\ \href@noop {} {Ph.D. thesis},\ \bibinfo  {school} {Massachusetts Institute of Technology.} (\bibinfo {year} {2006}{\natexlab{b}})\BibitemShut {NoStop}%
\bibitem [{\citenamefont {Thrane}\ \emph {et~al.}(2009)\citenamefont {Thrane}, \citenamefont {Ballmer}, \citenamefont {Romano}, \citenamefont {Mitra}, \citenamefont {Talukder}, \citenamefont {Bose},\ and\ \citenamefont {Mandic}}]{Sph_Thrane}%
  \BibitemOpen
  \bibfield  {author} {\bibinfo {author} {\bibfnamefont {E.}~\bibnamefont {Thrane}}, \bibinfo {author} {\bibfnamefont {S.}~\bibnamefont {Ballmer}}, \bibinfo {author} {\bibfnamefont {J.~D.}\ \bibnamefont {Romano}}, \bibinfo {author} {\bibfnamefont {S.}~\bibnamefont {Mitra}}, \bibinfo {author} {\bibfnamefont {D.}~\bibnamefont {Talukder}}, \bibinfo {author} {\bibfnamefont {S.}~\bibnamefont {Bose}}, \ and\ \bibinfo {author} {\bibfnamefont {V.}~\bibnamefont {Mandic}},\ }\href {\doibase 10.1103/physrevd.80.122002} {\bibfield  {journal} {\bibinfo  {journal} {Physical Review D}\ }\textbf {\bibinfo {volume} {80}} (\bibinfo {year} {2009}),\ 10.1103/physrevd.80.122002}\BibitemShut {NoStop}%
\bibitem [{\citenamefont {Tsukada}\ \emph {et~al.}(2023)\citenamefont {Tsukada}, \citenamefont {Jaraba}, \citenamefont {Agarwal},\ and\ \citenamefont {Floden}}]{Tsukada2023}%
  \BibitemOpen
  \bibfield  {author} {\bibinfo {author} {\bibfnamefont {L.}~\bibnamefont {Tsukada}}, \bibinfo {author} {\bibfnamefont {S.}~\bibnamefont {Jaraba}}, \bibinfo {author} {\bibfnamefont {D.}~\bibnamefont {Agarwal}}, \ and\ \bibinfo {author} {\bibfnamefont {E.}~\bibnamefont {Floden}},\ }\href {\doibase 10.1103/physrevd.107.023024} {\bibfield  {journal} {\bibinfo  {journal} {Physical Review D}\ }\textbf {\bibinfo {volume} {107}} (\bibinfo {year} {2023}),\ 10.1103/physrevd.107.023024}\BibitemShut {NoStop}%
\bibitem [{\citenamefont {Parida}\ \emph {et~al.}(2016)\citenamefont {Parida}, \citenamefont {Mitra},\ and\ \citenamefont {Jhingan}}]{Parida_2016}%
  \BibitemOpen
  \bibfield  {author} {\bibinfo {author} {\bibfnamefont {A.}~\bibnamefont {Parida}}, \bibinfo {author} {\bibfnamefont {S.}~\bibnamefont {Mitra}}, \ and\ \bibinfo {author} {\bibfnamefont {S.}~\bibnamefont {Jhingan}},\ }\href {\doibase 10.1088/1475-7516/2016/04/024} {\bibfield  {journal} {\bibinfo  {journal} {Journal of Cosmology and Astroparticle Physics}\ }\textbf {\bibinfo {volume} {2016}},\ \bibinfo {pages} {024–024} (\bibinfo {year} {2016})}\BibitemShut {NoStop}%
\bibitem [{\citenamefont {Boileau}\ \emph {et~al.}(2021)\citenamefont {Boileau}, \citenamefont {Lamberts}, \citenamefont {Christensen}, \citenamefont {Cornish},\ and\ \citenamefont {Meyer}}]{Boileau_2021}%
  \BibitemOpen
  \bibfield  {author} {\bibinfo {author} {\bibfnamefont {G.}~\bibnamefont {Boileau}}, \bibinfo {author} {\bibfnamefont {A.}~\bibnamefont {Lamberts}}, \bibinfo {author} {\bibfnamefont {N.}~\bibnamefont {Christensen}}, \bibinfo {author} {\bibfnamefont {N.~J.}\ \bibnamefont {Cornish}}, \ and\ \bibinfo {author} {\bibfnamefont {R.}~\bibnamefont {Meyer}},\ }\href {\doibase 10.1093/mnras/stab2575} {\bibfield  {journal} {\bibinfo  {journal} {Monthly Notices of the Royal Astronomical Society}\ }\textbf {\bibinfo {volume} {508}},\ \bibinfo {pages} {803–826} (\bibinfo {year} {2021})}\BibitemShut {NoStop}%
\bibitem [{\citenamefont {Suresh}\ \emph {et~al.}(2021)\citenamefont {Suresh}, \citenamefont {Agarwal},\ and\ \citenamefont {Mitra}}]{Suresh_2021}%
  \BibitemOpen
  \bibfield  {author} {\bibinfo {author} {\bibfnamefont {J.}~\bibnamefont {Suresh}}, \bibinfo {author} {\bibfnamefont {D.}~\bibnamefont {Agarwal}}, \ and\ \bibinfo {author} {\bibfnamefont {S.}~\bibnamefont {Mitra}},\ }\href {\doibase 10.1103/physrevd.104.102003} {\bibfield  {journal} {\bibinfo  {journal} {Physical Review D}\ }\textbf {\bibinfo {volume} {104}} (\bibinfo {year} {2021}),\ 10.1103/physrevd.104.102003}\BibitemShut {NoStop}%
\bibitem [{\citenamefont {Lillo}\ and\ \citenamefont {Suresh}(2024)}]{Federico_2024}%
  \BibitemOpen
  \bibfield  {author} {\bibinfo {author} {\bibfnamefont {F.~D.}\ \bibnamefont {Lillo}}\ and\ \bibinfo {author} {\bibfnamefont {J.}~\bibnamefont {Suresh}},\ }\href {https://arxiv.org/abs/2310.05823} {\enquote {\bibinfo {title} {Estimating astrophysical population properties using a multi-component stochastic gravitational-wave background search},}\ } (\bibinfo {year} {2024}),\ \Eprint {http://arxiv.org/abs/2310.05823} {arXiv:2310.05823 [gr-qc]} \BibitemShut {NoStop}%
\bibitem [{\citenamefont {Tsukada}(2023)}]{Tsukada2023nonGR}%
  \BibitemOpen
  \bibfield  {author} {\bibinfo {author} {\bibfnamefont {L.}~\bibnamefont {Tsukada}},\ }\href@noop {} {\enquote {\bibinfo {title} {Extension of the bayesian searches for anisotropic stochastic gravitational-wave background with non-tensorial polarizations},}\ } (\bibinfo {year} {2023}),\ \Eprint {http://arxiv.org/abs/2308.09020} {arXiv:2308.09020 [astro-ph.IM]} \BibitemShut {NoStop}%
\bibitem [{\citenamefont {Allen}\ and\ \citenamefont {Ottewill}(1997)}]{Allen_ORF}%
  \BibitemOpen
  \bibfield  {author} {\bibinfo {author} {\bibfnamefont {B.}~\bibnamefont {Allen}}\ and\ \bibinfo {author} {\bibfnamefont {A.~C.}\ \bibnamefont {Ottewill}},\ }\href {\doibase 10.1103/physrevd.56.545} {\bibfield  {journal} {\bibinfo  {journal} {Physical Review D}\ }\textbf {\bibinfo {volume} {56}},\ \bibinfo {pages} {545–563} (\bibinfo {year} {1997})}\BibitemShut {NoStop}%
\bibitem [{\citenamefont {Romano}\ and\ \citenamefont {Cornish}(2017)}]{Romano_2017}%
  \BibitemOpen
  \bibfield  {author} {\bibinfo {author} {\bibfnamefont {J.~D.}\ \bibnamefont {Romano}}\ and\ \bibinfo {author} {\bibfnamefont {N.~J.}\ \bibnamefont {Cornish}},\ }\href {\doibase 10.1007/s41114-017-0004-1} {\bibfield  {journal} {\bibinfo  {journal} {Living Reviews in Relativity}\ }\textbf {\bibinfo {volume} {20}} (\bibinfo {year} {2017}),\ 10.1007/s41114-017-0004-1}\BibitemShut {NoStop}%
\bibitem [{\citenamefont {Abbott}\ \emph {et~al.}(2017{\natexlab{c}})\citenamefont {Abbott} \emph {et~al.}}]{O1Isotropic}%
  \BibitemOpen
  \bibfield  {author} {\bibinfo {author} {\bibfnamefont {B.~P.}\ \bibnamefont {Abbott}} \emph {et~al.} (\bibinfo {collaboration} {LIGO Scientific Collaboration and Virgo Collaboration}),\ }\href {\doibase 10.1103/PhysRevLett.118.121101} {\bibfield  {journal} {\bibinfo  {journal} {Phys. Rev. Lett.}\ }\textbf {\bibinfo {volume} {118}},\ \bibinfo {pages} {121101} (\bibinfo {year} {2017}{\natexlab{c}})}\BibitemShut {NoStop}%
\bibitem [{\citenamefont {Abbott}\ \emph {et~al.}(2019)\citenamefont {Abbott} \emph {et~al.}}]{O2Isotropic}%
  \BibitemOpen
  \bibfield  {author} {\bibinfo {author} {\bibfnamefont {B.~P.}\ \bibnamefont {Abbott}} \emph {et~al.} (\bibinfo {collaboration} {LIGO Scientific and Virgo Collaboration}),\ }\href {\doibase 10.1103/PhysRevD.100.061101} {\bibfield  {journal} {\bibinfo  {journal} {Phys. Rev. D}\ }\textbf {\bibinfo {volume} {100}},\ \bibinfo {pages} {061101} (\bibinfo {year} {2019})}\BibitemShut {NoStop}%
\bibitem [{\citenamefont {Ain}\ \emph {et~al.}(2015)\citenamefont {Ain}, \citenamefont {Dalvi},\ and\ \citenamefont {Mitra}}]{Ain_2015}%
  \BibitemOpen
  \bibfield  {author} {\bibinfo {author} {\bibfnamefont {A.}~\bibnamefont {Ain}}, \bibinfo {author} {\bibfnamefont {P.}~\bibnamefont {Dalvi}}, \ and\ \bibinfo {author} {\bibfnamefont {S.}~\bibnamefont {Mitra}},\ }\href {\doibase 10.1103/physrevd.92.022003} {\bibfield  {journal} {\bibinfo  {journal} {Physical Review D}\ }\textbf {\bibinfo {volume} {92}} (\bibinfo {year} {2015}),\ 10.1103/physrevd.92.022003}\BibitemShut {NoStop}%
\bibitem [{\citenamefont {Allen}\ and\ \citenamefont {Romano}(1999)}]{Allen_1999}%
  \BibitemOpen
  \bibfield  {author} {\bibinfo {author} {\bibfnamefont {B.}~\bibnamefont {Allen}}\ and\ \bibinfo {author} {\bibfnamefont {J.~D.}\ \bibnamefont {Romano}},\ }\href {\doibase 10.1103/PhysRevD.59.102001} {\bibfield  {journal} {\bibinfo  {journal} {Phys. Rev. D}\ }\textbf {\bibinfo {volume} {59}},\ \bibinfo {pages} {102001} (\bibinfo {year} {1999})}\BibitemShut {NoStop}%
\bibitem [{\citenamefont {Ashton}\ \emph {et~al.}(2019)\citenamefont {Ashton}, \citenamefont {Hübner}, \citenamefont {Lasky}, \citenamefont {Talbot}, \citenamefont {Ackley}, \citenamefont {Biscoveanu}, \citenamefont {Chu}, \citenamefont {Divakarla}, \citenamefont {Easter}, \citenamefont {Goncharov}, \citenamefont {Vivanco}, \citenamefont {Harms}, \citenamefont {Lower}, \citenamefont {Meadors}, \citenamefont {Melchor}, \citenamefont {Payne}, \citenamefont {Pitkin}, \citenamefont {Powell}, \citenamefont {Sarin}, \citenamefont {Smith},\ and\ \citenamefont {Thrane}}]{bilby_2019}%
  \BibitemOpen
  \bibfield  {author} {\bibinfo {author} {\bibfnamefont {G.}~\bibnamefont {Ashton}}, \bibinfo {author} {\bibfnamefont {M.}~\bibnamefont {Hübner}}, \bibinfo {author} {\bibfnamefont {P.~D.}\ \bibnamefont {Lasky}}, \bibinfo {author} {\bibfnamefont {C.}~\bibnamefont {Talbot}}, \bibinfo {author} {\bibfnamefont {K.}~\bibnamefont {Ackley}}, \bibinfo {author} {\bibfnamefont {S.}~\bibnamefont {Biscoveanu}}, \bibinfo {author} {\bibfnamefont {Q.}~\bibnamefont {Chu}}, \bibinfo {author} {\bibfnamefont {A.}~\bibnamefont {Divakarla}}, \bibinfo {author} {\bibfnamefont {P.~J.}\ \bibnamefont {Easter}}, \bibinfo {author} {\bibfnamefont {B.}~\bibnamefont {Goncharov}}, \bibinfo {author} {\bibfnamefont {F.~H.}\ \bibnamefont {Vivanco}}, \bibinfo {author} {\bibfnamefont {J.}~\bibnamefont {Harms}}, \bibinfo {author} {\bibfnamefont {M.~E.}\ \bibnamefont {Lower}}, \bibinfo {author} {\bibfnamefont {G.~D.}\ \bibnamefont {Meadors}}, \bibinfo {author} {\bibfnamefont {D.}~\bibnamefont {Melchor}}, \bibinfo {author} {\bibfnamefont
  {E.}~\bibnamefont {Payne}}, \bibinfo {author} {\bibfnamefont {M.~D.}\ \bibnamefont {Pitkin}}, \bibinfo {author} {\bibfnamefont {J.}~\bibnamefont {Powell}}, \bibinfo {author} {\bibfnamefont {N.}~\bibnamefont {Sarin}}, \bibinfo {author} {\bibfnamefont {R.~J.~E.}\ \bibnamefont {Smith}}, \ and\ \bibinfo {author} {\bibfnamefont {E.}~\bibnamefont {Thrane}},\ }\href {\doibase 10.3847/1538-4365/ab06fc} {\bibfield  {journal} {\bibinfo  {journal} {The Astrophysical Journal Supplement Series}\ }\textbf {\bibinfo {volume} {241}},\ \bibinfo {pages} {27} (\bibinfo {year} {2019})}\BibitemShut {NoStop}%
\bibitem [{\citenamefont {Romero-Shaw}\ \emph {et~al.}(2020)\citenamefont {Romero-Shaw}, \citenamefont {Talbot}, \citenamefont {Biscoveanu}, \citenamefont {D’Emilio}, \citenamefont {Ashton}, \citenamefont {Berry}, \citenamefont {Coughlin}, \citenamefont {Galaudage}, \citenamefont {Hoy}, \citenamefont {Hübner}, \citenamefont {Phukon}, \citenamefont {Pitkin}, \citenamefont {Rizzo}, \citenamefont {Sarin}, \citenamefont {Smith}, \citenamefont {Stevenson}, \citenamefont {Vajpeyi}, \citenamefont {Arène}, \citenamefont {Athar}, \citenamefont {Banagiri}, \citenamefont {Bose}, \citenamefont {Carney}, \citenamefont {Chatziioannou}, \citenamefont {Clark}, \citenamefont {Colleoni}, \citenamefont {Cotesta}, \citenamefont {Edelman}, \citenamefont {Estellés}, \citenamefont {García-Quirós}, \citenamefont {Ghosh}, \citenamefont {Green}, \citenamefont {Haster}, \citenamefont {Husa}, \citenamefont {Keitel}, \citenamefont {Kim}, \citenamefont {Hernandez-Vivanco}, \citenamefont {Magaña Hernandez}, \citenamefont
  {Karathanasis}, \citenamefont {Lasky}, \citenamefont {De Lillo}, \citenamefont {Lower}, \citenamefont {Macleod}, \citenamefont {Mateu-Lucena}, \citenamefont {Miller}, \citenamefont {Millhouse}, \citenamefont {Morisaki}, \citenamefont {Oh}, \citenamefont {Ossokine}, \citenamefont {Payne}, \citenamefont {Powell}, \citenamefont {Pratten}, \citenamefont {Pürrer}, \citenamefont {Ramos-Buades}, \citenamefont {Raymond}, \citenamefont {Thrane}, \citenamefont {Veitch}, \citenamefont {Williams}, \citenamefont {Williams},\ and\ \citenamefont {Xiao}}]{bilby_2020}%
  \BibitemOpen
  \bibfield  {author} {\bibinfo {author} {\bibfnamefont {I.~M.}\ \bibnamefont {Romero-Shaw}}, \bibinfo {author} {\bibfnamefont {C.}~\bibnamefont {Talbot}}, \bibinfo {author} {\bibfnamefont {S.}~\bibnamefont {Biscoveanu}}, \bibinfo {author} {\bibfnamefont {V.}~\bibnamefont {D’Emilio}}, \bibinfo {author} {\bibfnamefont {G.}~\bibnamefont {Ashton}}, \bibinfo {author} {\bibfnamefont {C.~P.~L.}\ \bibnamefont {Berry}}, \bibinfo {author} {\bibfnamefont {S.}~\bibnamefont {Coughlin}}, \bibinfo {author} {\bibfnamefont {S.}~\bibnamefont {Galaudage}}, \bibinfo {author} {\bibfnamefont {C.}~\bibnamefont {Hoy}}, \bibinfo {author} {\bibfnamefont {M.}~\bibnamefont {Hübner}}, \bibinfo {author} {\bibfnamefont {K.~S.}\ \bibnamefont {Phukon}}, \bibinfo {author} {\bibfnamefont {M.}~\bibnamefont {Pitkin}}, \bibinfo {author} {\bibfnamefont {M.}~\bibnamefont {Rizzo}}, \bibinfo {author} {\bibfnamefont {N.}~\bibnamefont {Sarin}}, \bibinfo {author} {\bibfnamefont {R.}~\bibnamefont {Smith}}, \bibinfo {author} {\bibfnamefont
  {S.}~\bibnamefont {Stevenson}}, \bibinfo {author} {\bibfnamefont {A.}~\bibnamefont {Vajpeyi}}, \bibinfo {author} {\bibfnamefont {M.}~\bibnamefont {Arène}}, \bibinfo {author} {\bibfnamefont {K.}~\bibnamefont {Athar}}, \bibinfo {author} {\bibfnamefont {S.}~\bibnamefont {Banagiri}}, \bibinfo {author} {\bibfnamefont {N.}~\bibnamefont {Bose}}, \bibinfo {author} {\bibfnamefont {M.}~\bibnamefont {Carney}}, \bibinfo {author} {\bibfnamefont {K.}~\bibnamefont {Chatziioannou}}, \bibinfo {author} {\bibfnamefont {J.~A.}\ \bibnamefont {Clark}}, \bibinfo {author} {\bibfnamefont {M.}~\bibnamefont {Colleoni}}, \bibinfo {author} {\bibfnamefont {R.}~\bibnamefont {Cotesta}}, \bibinfo {author} {\bibfnamefont {B.}~\bibnamefont {Edelman}}, \bibinfo {author} {\bibfnamefont {H.}~\bibnamefont {Estellés}}, \bibinfo {author} {\bibfnamefont {C.}~\bibnamefont {García-Quirós}}, \bibinfo {author} {\bibfnamefont {A.}~\bibnamefont {Ghosh}}, \bibinfo {author} {\bibfnamefont {R.}~\bibnamefont {Green}}, \bibinfo {author} {\bibfnamefont
  {C.-J.}\ \bibnamefont {Haster}}, \bibinfo {author} {\bibfnamefont {S.}~\bibnamefont {Husa}}, \bibinfo {author} {\bibfnamefont {D.}~\bibnamefont {Keitel}}, \bibinfo {author} {\bibfnamefont {A.~X.}\ \bibnamefont {Kim}}, \bibinfo {author} {\bibfnamefont {F.}~\bibnamefont {Hernandez-Vivanco}}, \bibinfo {author} {\bibfnamefont {I.}~\bibnamefont {Magaña Hernandez}}, \bibinfo {author} {\bibfnamefont {C.}~\bibnamefont {Karathanasis}}, \bibinfo {author} {\bibfnamefont {P.~D.}\ \bibnamefont {Lasky}}, \bibinfo {author} {\bibfnamefont {N.}~\bibnamefont {De Lillo}}, \bibinfo {author} {\bibfnamefont {M.~E.}\ \bibnamefont {Lower}}, \bibinfo {author} {\bibfnamefont {D.}~\bibnamefont {Macleod}}, \bibinfo {author} {\bibfnamefont {M.}~\bibnamefont {Mateu-Lucena}}, \bibinfo {author} {\bibfnamefont {A.}~\bibnamefont {Miller}}, \bibinfo {author} {\bibfnamefont {M.}~\bibnamefont {Millhouse}}, \bibinfo {author} {\bibfnamefont {S.}~\bibnamefont {Morisaki}}, \bibinfo {author} {\bibfnamefont {S.~H.}\ \bibnamefont {Oh}}, \bibinfo
  {author} {\bibfnamefont {S.}~\bibnamefont {Ossokine}}, \bibinfo {author} {\bibfnamefont {E.}~\bibnamefont {Payne}}, \bibinfo {author} {\bibfnamefont {J.}~\bibnamefont {Powell}}, \bibinfo {author} {\bibfnamefont {G.}~\bibnamefont {Pratten}}, \bibinfo {author} {\bibfnamefont {M.}~\bibnamefont {Pürrer}}, \bibinfo {author} {\bibfnamefont {A.}~\bibnamefont {Ramos-Buades}}, \bibinfo {author} {\bibfnamefont {V.}~\bibnamefont {Raymond}}, \bibinfo {author} {\bibfnamefont {E.}~\bibnamefont {Thrane}}, \bibinfo {author} {\bibfnamefont {J.}~\bibnamefont {Veitch}}, \bibinfo {author} {\bibfnamefont {D.}~\bibnamefont {Williams}}, \bibinfo {author} {\bibfnamefont {M.~J.}\ \bibnamefont {Williams}}, \ and\ \bibinfo {author} {\bibfnamefont {L.}~\bibnamefont {Xiao}},\ }\href {\doibase 10.1093/mnras/staa2850} {\bibfield  {journal} {\bibinfo  {journal} {Monthly Notices of the Royal Astronomical Society}\ }\textbf {\bibinfo {volume} {499}},\ \bibinfo {pages} {3295–3319} (\bibinfo {year} {2020})}\BibitemShut {NoStop}%
\bibitem [{\citenamefont {Speagle}(2020)}]{dynesty}%
  \BibitemOpen
  \bibfield  {author} {\bibinfo {author} {\bibfnamefont {J.~S.}\ \bibnamefont {Speagle}},\ }\href {\doibase 10.1093/mnras/staa278} {\bibfield  {journal} {\bibinfo  {journal} {Monthly Notices of the Royal Astronomical Society}\ }\textbf {\bibinfo {volume} {493}},\ \bibinfo {pages} {3132–3158} (\bibinfo {year} {2020})}\BibitemShut {NoStop}%
\bibitem [{\citenamefont {Abbott}\ and\ \citenamefont {others"}(2020)}]{Prospects_forALIGO}%
  \BibitemOpen
  \bibfield  {author} {\bibinfo {author} {\bibfnamefont {B.~P.}\ \bibnamefont {Abbott}}\ and\ \bibinfo {author} {\bibnamefont {others"}},\ }\href {\doibase 10.1007/s41114-020-00026-9} {\bibfield  {journal} {\bibinfo  {journal} {Living Reviews in Relativity}\ }\textbf {\bibinfo {volume} {23}} (\bibinfo {year} {2020}),\ 10.1007/s41114-020-00026-9}\BibitemShut {NoStop}%
\bibitem [{\citenamefont {Callister}\ \emph {et~al.}(2017)\citenamefont {Callister}, \citenamefont {Biscoveanu}, \citenamefont {Christensen}, \citenamefont {Isi}, \citenamefont {Matas}, \citenamefont {Minazzoli}, \citenamefont {Regimbau}, \citenamefont {Sakellariadou}, \citenamefont {Tasson},\ and\ \citenamefont {Thrane}}]{Callister_2017}%
  \BibitemOpen
  \bibfield  {author} {\bibinfo {author} {\bibfnamefont {T.}~\bibnamefont {Callister}}, \bibinfo {author} {\bibfnamefont {A.~S.}\ \bibnamefont {Biscoveanu}}, \bibinfo {author} {\bibfnamefont {N.}~\bibnamefont {Christensen}}, \bibinfo {author} {\bibfnamefont {M.}~\bibnamefont {Isi}}, \bibinfo {author} {\bibfnamefont {A.}~\bibnamefont {Matas}}, \bibinfo {author} {\bibfnamefont {O.}~\bibnamefont {Minazzoli}}, \bibinfo {author} {\bibfnamefont {T.}~\bibnamefont {Regimbau}}, \bibinfo {author} {\bibfnamefont {M.}~\bibnamefont {Sakellariadou}}, \bibinfo {author} {\bibfnamefont {J.}~\bibnamefont {Tasson}}, \ and\ \bibinfo {author} {\bibfnamefont {E.}~\bibnamefont {Thrane}},\ }\href {\doibase 10.1103/physrevx.7.041058} {\bibfield  {journal} {\bibinfo  {journal} {Physical Review X}\ }\textbf {\bibinfo {volume} {7}} (\bibinfo {year} {2017}),\ 10.1103/physrevx.7.041058}\BibitemShut {NoStop}%
\bibitem [{\citenamefont {Kass}\ and\ \citenamefont {Raftery}(1995)}]{BFcriteria}%
  \BibitemOpen
  \bibfield  {author} {\bibinfo {author} {\bibfnamefont {R.~E.}\ \bibnamefont {Kass}}\ and\ \bibinfo {author} {\bibfnamefont {A.~E.}\ \bibnamefont {Raftery}},\ }\href {\doibase 10.1080/01621459.1995.10476572} {\bibfield  {journal} {\bibinfo  {journal} {Journal of the American Statistical Association}\ }\textbf {\bibinfo {volume} {90}},\ \bibinfo {pages} {773} (\bibinfo {year} {1995})}\BibitemShut {NoStop}%
\end{thebibliography}%
\end{document}